\documentclass[twocolumn,english,aps,reprint]{revtex4}
\usepackage[T1]{fontenc}
\usepackage[latin9]{inputenc}
\usepackage[active]{srcltx}
\usepackage{amstext}
\usepackage{amsmath}
\usepackage{amssymb}
\usepackage{graphicx}
\usepackage{color}
\usepackage{esint}
\usepackage{bbold}
\usepackage{dsfont}
\usepackage{mathtools}
\usepackage{changes}

\makeatletter
\@ifundefined{textcolor}{}
{%
 \definecolor{BLACK}{gray}{0}
 \definecolor{WHITE}{gray}{1}
 \definecolor{RED}{rgb}{1,0,0}
 \definecolor{GREEN}{rgb}{0,1,0}
 \definecolor{BLUE}{rgb}{0,0,1}
 \definecolor{CYAN}{cmyk}{1,0,0,0}
 \definecolor{MAGENTA}{cmyk}{0,1,0,0}
 \definecolor{YELLOW}{cmyk}{0,0,1,0}
}

\providecommand{\cf}[1]{#1}

\makeatother

\usepackage{babel}
\begin{document}

\preprint{This line only printed with preprint option}

\title{Electron-magnon scattering in elementary ferromagnets from first
principles: lifetime broadening and \cf{band anomalies}}


\author{Mathias C. T. D. M\"uller, Stefan Bl\"ugel, and Christoph Friedrich}


\affiliation{Peter Gr\"unberg Institut and Institute for Advanced Simulation,
Forschungszentrum J\"ulich, 52425 J\"ulich, Germany}
\begin{abstract}
We study the electron-magnon scattering in
bulk Fe, Co, and Ni within the framework of many-body perturbation
theory implemented in the full-potential linearized augmented-plane-wave 
method.
To this end, a $\mathbf{k}$-dependent self-energy ($GT$ self-energy) describing the scattering of electrons and magnons
is constructed from the solution of
a Bethe-Salpeter equation for the two-particle (electron-hole) Green function,
in which single-particle
Stoner and collective spin-wave excitations (magnons) are treated 
on the same footing.
Partial self-consistency is achieved by the alignment of the chemical potentials.
The resulting renormalized
electronic band structures exhibit strong spin-dependent lifetime effects close to
the Fermi energy, which are strongest in Fe. 
The renormalization can give rise to a loss of quasiparticle character close to the Fermi energy, 
which we attribute to electron scattering with spatially extended spin waves.
\cf{This scattering is also responsible for dispersion anomalies in conduction
bands of iron and for the formation of satellite bands in nickel.}
Furthermore, we find a band anomaly at a binding energy of 1.5~eV in iron, which results 
from a coupling of the quasihole with single-particle excitations that form a peak in the Stoner
continuum. This band anomaly was recently observed in photoemission experiments. \cf{On the theory side, we show that the contribution of the
Goldstone mode to the $GT$ self-energy is expected to (nearly) vanish in the
long-wavelength limit. We also present an in-depth discussion about
the possible violation of causality when an incomplete subset of
self-energy diagrams is chosen.}
\end{abstract}
\maketitle

\section{Introduction}
The interaction of electrons and spin excitations plays a fundamental role for a 
wide variety of phenomena. For example, 
spin-polarized currents 
depolarize due to their interaction with magnons~\cite{zutic04}, the characteristic temperature dependence 
of the tunneling magnetoresistance (TMR) \cite{macdonald98} is determined by the electron 
scattering by magnons, and in nanospintronics 
the spin and charge currents that flow through nanostructures 
can be strongly affected by the electron-magnon interaction \cite{balashov06,balashov08,schweflinghaus14,schweflinghaus16}. 
Moreover, it is speculated that the electron-magnon interaction is the 
origin of the superconductivity in Fe pnictides \cite{dagotto94,scalapino95,sasmal08,dahm09}. 

The elementary ferromagnets Fe, Co, and Ni are suitable model systems to study the electron-magnon interaction. They 
form a class of intermediately correlated materials in which localized 
$d$ states close to the Fermi level are embedded in a free-electron-like band structure. While density-functional theory (DFT), 
employing either the local-spin-density (LSDA) or the generalized gradient approximation (GGA) to the exchange-correlation potential,
is able to capture the ground-state properties of these materials, it fails
to yield accurate excited-state properties.
For example, the exchange splitting of these materials is often
overestimated within DFT calculations.
Nickel is an extreme case in that the exchange splitting is almost a factor of two too large. 

A strongly spin-dependent band broadening is observed in
angle-resolved~\cite{monastra02} and two-photon photoemission experiments \cite{knorren00}, 
which indicates that the electron-magnon interaction
plays an important role in $3d$ ferromagnets.
In addition, recent angle-resolved photoemission spectroscopy (ARPES) experiments reveal pronounced differences 
in the quasiparticle dispersion compared to DFT. 
ARPES measurements, for example, have exhibited anomalous kinks in the band
dispersion of iron surface states \cite{schaefer04} and nickel bulk states
\cite{hofmann09}. These kinks appear at binding energies much larger than
what one would expect for a phonon-mediated band renormalization. It has
been speculated that these band anomalies are a footprint of electron-magnon
scattering because the energetic position of the kinks corresponds to typical
magnon energies in the materials. We will later show that a similar band 
anomaly in iron, which was observed in ARPES spectra \cite{Mlynczak19} at a comparatively large binding energy of 1.5 eV, can indeed be explained within our theory. 
(In this particular case, the theoretical prediction predated the
experimental observation.)

%
%
%


The appearance of kinks in the electronic band dispersion and the strong
band broadening of the quasiparticle peaks clearly go beyond the scope of
a static mean-field theory such as DFT and calls for a genuine many-body
description. 
One such method is the dynamical mean-field theory (DMFT), 
which maps the interacting many-body system onto an Anderson impurity
model~\cite{georges96,kotliar06}. DMFT relies on the choice of model
parameters, the effective intra-atomic interaction parameters $U$ and $J$ as
well as the intra- and inter-atomic hopping parameters.
In applications to real materials, the
latter are usually taken from a L(S)DA mean-field solution. 
An early implementation~\cite{katsnelson99} of this LSDA+DMFT method
showed that majority hole states are strongly damped in iron.
This result has been confirmed by Grechnev~\textit{et al.}~\cite{grechnev07} who 
studied Fe, Co, and Ni with the same approach.
They found 
a strong damping of the majority quasiparticle states and, in addition, a
shallow satellite feature below the bottom of the $d$ bands. 
A comparison of ARPES spectra to LSDA+DMFT calculations~\cite{sanchez-barriga09,sanchez-barriga10,sanchez-barriga12}
revealed that 
the agreement between experiment and theory is considerably improved with
respect to LSDA, but still the linewidths and the effective masses tend 
to be underestimated compared to experiment. 
So far, no evidence of band dispersion anomalies has been found in the
theoretical studies.

While allowing for an efficient treatment of many-body effects, the DMFT
method suffers from some badly controlled approximations.
Above all, the usage
of the impurity model essentially amounts to neglecting the momentum
dependence of the many-body scattering processes. Furthermore, the choice of
the model parameters introduces an element of arbitrariness. Often they are
fitted to experiment, which limits the predictive power of the method. 
Finally, since LSDA already contains the electron-electron
interaction in an approximate way, it is necessary to apply a double-counting
correction, for which no unique definition exists.

In this work, we employ an alternative description that avoids the usage of a
model. The single-particle wave functions and propagators are allowed to
extend over the whole (infinite) crystal. In this way, the
momentum dependence is retained, and there is no need for model parameters
nor a double-counting correction.
Many-body perturbation theory (MBPT) is employed to construct 
an approximation to the electronic self-energy
$\Sigma(\mathbf{r},\mathbf{r}';\omega)$. This effective 
scattering potential describes many-body exchange and correlation scattering processes 
that an electron or hole experiences as it propagates through a many-electron system. 
In other words, the self-energy connects the non-interacting mean-field
system to the real interacting system. 
The solution of the Dyson equation then yields the single-particle Green
function of the interacting system, and the imaginary part of which is
directly related to the photoemission spectra~\cite{onida02}.

The most popular approximation to the electronic self-energy is the $GW$
approximation, which has been shown to yield
accurate band structures for a wide range of materials. For example, it is known 
to have a strong effect on the band gaps of semiconductors and insulators
\cite{Aulbur99}, which are corrected from their (usually underestimated)
DFT values towards experiment. The $GW$ method has also been
applied to the elementary ferromagnets, where it partly cures the
shortcomings of LSDA.
While the $3d$ band width is typically overestimated in LSDA, the $GW$ approximation applied 
to iron~\cite{yamasaki03} and nickel~\cite{aryasetiawan92,friedrich10}
yields results in better agreement to the experimental values~\cite{eastman80,hoechst77,himpsel79,eberhardt80,heimann81,kirby85,hoechst76,hoechst76a}.
The explicit electron-electron scattering of the $GW$ self-energy gives rise to a lifetime
broadening of the bands, which is, however, too small to explain the
broadening seen in photoemission experiments, indicating that the $GW$ self-energy misses some
important scattering processes in these materials.

In this work, we derive from the Hedin equations~\cite{hedin65} a first-principles self-energy approximation that
describes the scattering of electrons and magnons.
The Hedin equations are a set of integro-differential equations that, if 
solved self-consistently, would yield, in
principle, the exact self-energy for a many-electron system. A full
self-consistent solution is not possible in practice, but the Hedin
equations can be used to derive approximations to the self-energy. For
example, the $GW$ approximation results from a single cycle through the equations
starting from $\Sigma=0$. By iteration of the Hedin equations, one
systematically generates more and more higher-order self-energy diagrams. In this way, 
we identify and select those scattering diagrams that are relevant for the coupling of
electrons to many-body spin excitations. Then, summing these ladder diagrams
to all orders in the interaction yields the $GT$ approximation, where $T$
stands for the magnon propagator, which describes the correlated motion of an
electron-hole pair with opposite spins. It is shown that the lowest-order
diagram of the $T$ matrix is of third order in $W$, which renders
double-counting corrections with $GW$ unnecessary.
The $GT$ approximation has a similar
mathematical form as the $GW$ self-energy in that it is given by a product
of the single-particle Green function $G$ and the effective magnon propagator $T$.
The implementation is, however, complicated by the fact that this $T$ matrix is a 
four-point quantity: It depends on four points in space and results from a
solution of a Bethe-Salpeter equation.
A numerical implementation has been realized by employing a basis set of 
Wannier functions, which allow for an efficient truncation 
of the four-point quantity in real space.
\cf{It should be noted that the present first-principles approach is based on a
formulation at absolute zero, whereas DMFT approaches usually require the assumption
of a finite electronic temperature.}

The theoretical foundation of the $GT$ self-energy is sketched in
Sec.~\ref{sec::Theory}. \cf{Section \ref{sec:Goldstone} discusses the
self-energy contribution of electronic scattering with acoustic magnons (Goldstone mode) 
in the long-wavelength limit.}
The implementation of the $GT$ self-energy 
within the \textsc{spex} code \cite{friedrich10} is described in Sec.~\ref{sec::Implementation}. 
\cf{The important topic of the violation of causality is discussed in detail in
Sec.~\ref{sec:causality}.}
We analyze the many-body renormalization of the band structure of the bulk elementary ferromagnets iron, cobalt, and 
nickel in Sec.~\ref{sec::Calculations}. The coupling of electrons to spin excitations leads to a pronounced spin-dependent 
lifetime broadening of the quasiparticle states. The lifetime broadening, which is particularly 
strong in iron, can lead to a complete loss of the quasiparticle character
in certain energy regions of the electronic spectrum -- with energy gaps
appearing in electronic bands, effectively cutting them in two.
Furthermore, we find
a band anomaly at higher binding energy in iron, in agreement with a very
recent ARPES study. Section~\ref{sec::Conclusions} gives a summary. 

\section{Theory} \label{sec::Theory}

Our goal is to construct a self-energy
that describes the many-body renormalization due to the scattering of
electrons and magnons. Here, we should understand the term \emph{magnon} to
comprise not only the collective spin-wave excitations but also the
single-particle Stoner excitations, which provide a decoherence channel,
through which the spin waves acquire a finite lifetime.
In our formulation, the two types of spin excitations are
treated on the same footing. They are intimately coupled to each other 
(two sides of the same coin)
and cannot, as a matter of principle, be treated separately.
In this sense, the self-energy will describe
the dressing of a propagating particle (electron or hole) through the
creation and annihilation of collective spin waves and 
single-particle Stoner excitations alike. We note that the spin-orbit coupling is
neglected.


\begin{figure}
\includegraphics[width=1.\columnwidth]{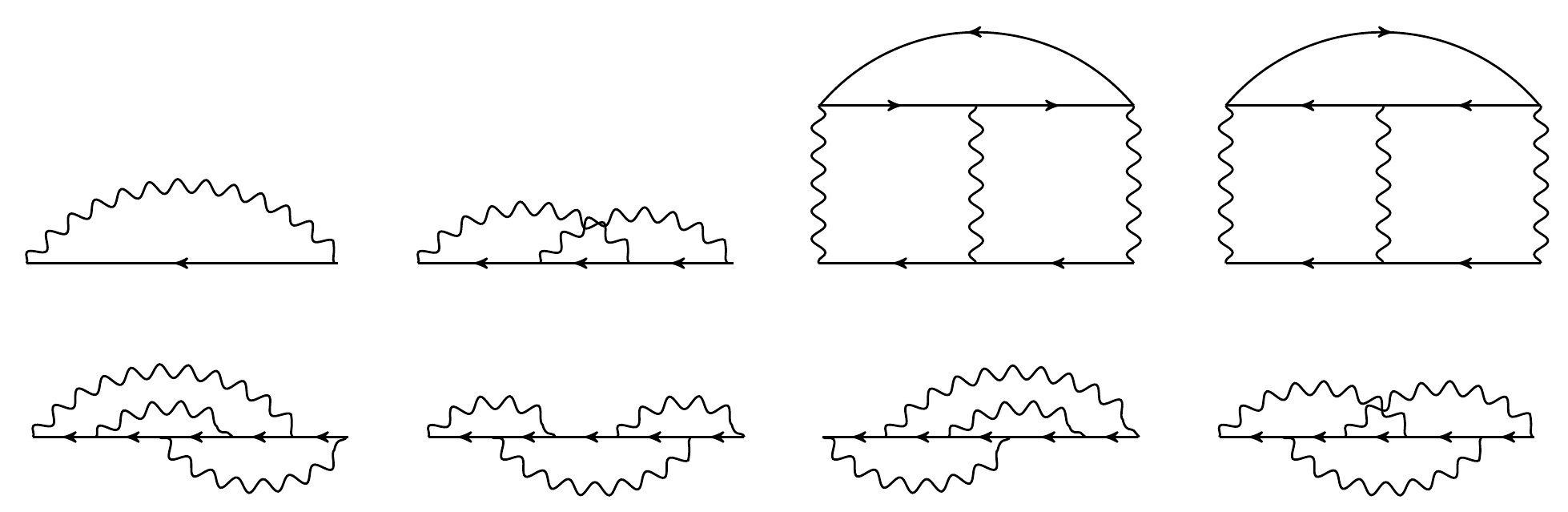}
\caption{\label{fig::SelfEnergyExpansion} Feynman diagrams of the 
Hedin self-energy expansion up to the third order in the screened
interaction $W$, which is denoted by 
a wiggly line. The Green function is denoted by an arrow.}
\end{figure} 

Figure \ref{fig::SelfEnergyExpansion} shows the self-energy expansion in
Feynman diagrams up to third order in the screened interaction $W$ as
derived from the Hedin equations. 
Cutting the expansion at the first order yields the $GW$ self-energy, 
an approximation commonly used in computational
condensed-matter physics. 
Despite its success, however,
it lacks a number of many-body scattering effects, for example, particle-particle
(electron-electron and hole-hole) and electron-hole scattering
as well as higher-order exchange processes.
To account for these, the $GW$ method has been combined
\cite{springer98,zhukov04,romaniello12} with the
$T$-matrix theory, in which the self-energy is expanded in terms of infinitely many ladder
diagrams, describing the correlated motion of two particles (electrons or
holes). This $GW$+$T$ approach is motivated phenomenologically: While $GW$
describes the correlation of itinerant $s$ and $p$ states, the ladder
diagrams account for correlation effects taking place in localized states.
However, as the $T$-matrix self-energy contains the Hartree ``tadpole'' diagram
and the direct term of the second Born approximation, which is also
contained in the $GW$ self-energy
\footnote{This is true if the diagrams are defined in terms of the free
Green function (excluding the electron-electron term of the Hamiltonian) and
the bare interaction or the effective Hubbard $U$ interaction. We will later
use the Kohn-Sham Green function and the fully screened interaction $W$, in which
case the two diagrams (omitted in the present formulation) would contain double-counting errors in themselves.},
a double-counting correction would be required in this
approach~\cite{springer98}.

Instead, we come back to the expansion shown in Fig.~\ref{fig::SelfEnergyExpansion}.
Apart from the $GW$ self-energy, this expansion consists of the second- and third-order
screened exchange diagrams
and two direct diagrams (the ones of third order in the first row) that also appear in the $T$-matrix approach.
These two diagrams have the characteristic form of ladder diagrams
where the rungs of the ladder correspond to the screened interaction
and the rails correspond to single-particle Green functions.
They describe the correlated motion of an electron-hole pair and a
particle pair (electron-electron or hole-hole), respectively.
If the electron and hole are of opposite spins, the corresponding
propagator can be seen as part of a solution of the Bethe-Salpeter equation
(BSE) for the transverse magnetic response
function~\cite{aryasetiawan99,sasioglu10,friedrich14,friedrich18}. 
The full BSE solution would comprise 
ladder diagrams of all orders, including the first and second as well as the
infinitely many higher-order diagrams. For example, the collective spin-wave
excitations arise from a resummation of ladder diagrams up to infinite order.
Transversal spin waves are low-energy excitations (going down to zero energy for
vanishing momentum in the absence of spin-orbit coupling). The
corresponding self-energy diagrams are therefore expected to yield the
principal low-energy scattering contribution in ferromagnets. 
The particle-particle diagrams, on the other hand, have been shown to be
responsible for the appearance of satellite features, e.g., the $6\,\textrm{eV}$ 
satellite in nickel~\cite{springer98}. The electron-hole
diagrams without a spin flip (electron and hole have the same spin) couple to density and
longitudinal spin fluctuations, which are of much higher energy (in the order of the plasma frequency) than the
transversal spin waves.
The other diagrams shown in Fig.~\ref{fig::SelfEnergyExpansion} cannot describe a coupling to spin fluctuations because,
in the absence of spin-orbit coupling, they have a single continuous Green-function line with a
unique spin quantum number. 


For these reasons, we have implemented the electron-hole self-energy ladder
diagrams, starting at the third order as prescribed by the Hedin
self-energy expansion. The summation is to all orders in $W$ 
(the third-order diagram generates the fourth-order diagram in the next
cycle of the Hedin equations and so on)
but excludes the first- and second-order diagrams, which avoids the
double-counting problem mentioned above. Iterating the Hedin
equations thus yields, in a natural way, a combination of the $GW$ and the
$T$-matrix self-energy that is free from double-counting errors.

\cf{At this point, we should mention that an implementation starting from the
third-order diagram was already considered in Ref.~\onlinecite{springer98}
but dismissed because the authors had realized that the renormalized Green
function violated a causality condition: The imaginary
part of the Green function showed an incorrect change of sign at large
(absolute)
energies. We observe the same behavior and discuss below in detail that
the sign change is not a numerical artifact but, in fact, expected to occur for the presently
chosen self-energy diagrams. However, we argue that this violation of
causality can be accepted as the sign change 
is observed in an energy region, which is not in the focus of the present
paper. In particular, it is expected that other self-energy diagrams,
neglected in the present work, will play a more important role there and should
eventually act to restore the correct sign.}

\begin{figure}
\includegraphics[width=0.85\columnwidth]{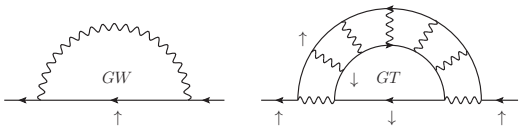}
\caption{\label{GW_GT}$GT$ diagram (right) drawn to resemble the $GW$
diagram (left). The $T$ matrix is a magnon propagator. The $GT$ self-energy
thus describes the emission and re-absorption of magnons in much the same
way as $GW$ describes the corresponding processes for plasmons.
Note that, after emission of the magnon, the spin of the propagating
particle has flipped, since the magnon carries a total spin of unity.
}
\end{figure}

From a physical point of view, the self-energy diagrams, which form
what we call the $GT$ self-energy in the following,
describe the emission and re-absorption of magnons. Here, the magnons should again be 
understood to comprise both spin waves and Stoner excitations. 
In this sense, the $T$ matrix can be viewed as an effective interaction
that acts through the exchange of magnons in much the same way as $W$ includes the
exchange of plasmons. To illustrate this analogy, we represent the self-energy diagram 
in a way that resembles the $GW$ diagram in Fig.~\ref{GW_GT}.
However, we will see that the $T$ matrix is a much
more complex quantity than $W$ as it is a four-point function, i.e., it depends on
four points in space (and time), whereas $W$ is a two-point function. 

The $GT$ self-energy is expected to yield an important
scattering channel for electrons in magnetic materials.
We should mention already at this point that for the present study we omit
the explicit evaluation of the $GW$ self-energy for simplicity and employ a corrected LSDA
solution instead. This solution is made to fulfill the Goldstone condition
for the spin excitations and has been shown~\cite{mueller16} to resemble the self-consistent solution obtained with the Coulomb-hole
screened-exchange (COHSEX) self-energy, the static limit of $GW$, \cf{and to
bring the exchange splittings closer to experiment.}




\begin{figure}
\includegraphics[width=1.\columnwidth]{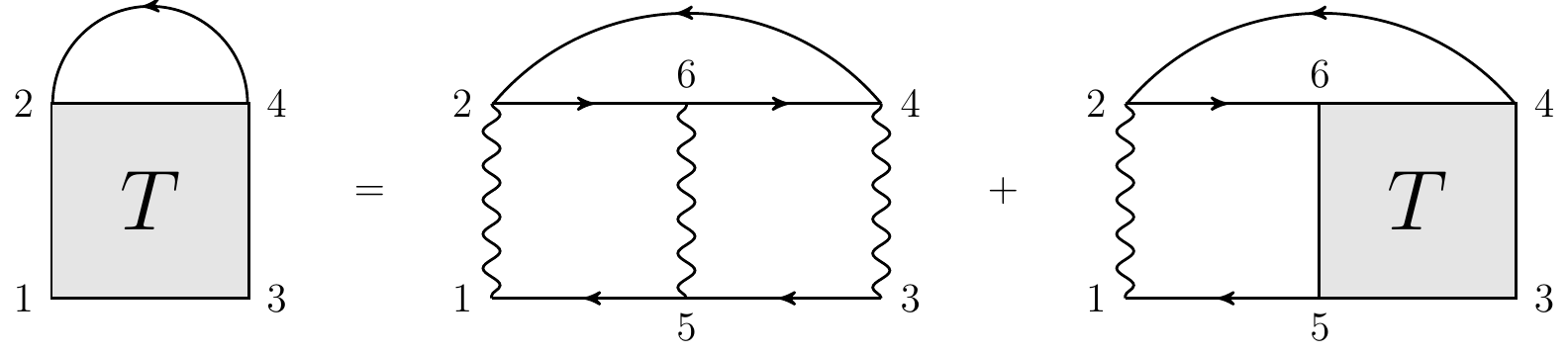}
\caption{\label{fig::GTSelfEnergy} Feynman diagram of the electron-magnon 
self-energy in the ladder approximation. Notation as in Fig.~\ref{fig::SelfEnergyExpansion}. 
It is assumed that the Green functions that are solely connected by the screened interaction have 
opposite spins.}
\end{figure} 

The $GT$ self-energy 
\begin{equation}\label{eq::ElMagSelfEnergy}
\Sigma^{\sigma}(13) = -\textrm{i} \int \textrm{d}2\,\textrm{d}4\,\,T^{\sigma\sigma^{\prime}}(12,34) G^{\sigma^{\prime}}(42)
\end{equation} 
with the spin indices $\sigma$ and $\sigma^\prime=-\sigma$ 
is written in terms of the interacting time-ordered single-particle Green function 
$G^{\sigma}$ and an electron-hole $T$ matrix. 
We have used the short-hand
notation $1=(\mathbf{r}_{1},t_{1})$, and atomic units are used throughout.
Figure \ref{fig::GTSelfEnergy}
shows a representation in Feynman diagrams.
In the ladder approximation, the multiple-scattering $T$ matrix is the
solution of the BSE
\begin{eqnarray}\label{eq::TmatrixBSE}
&&T^{\sigma\sigma^{\prime}}(12,34) = \int \textrm{d}5\,\textrm{d}6\,\, W(12) K^{\sigma\sigma^{\prime}}(12,56)
\nonumber\\
&&\times\left\{
W(56)K^{\sigma\sigma^{\prime}}(56,34) W(34) +
T^{\sigma\sigma^{\prime}}(56,34)
\right\}
\end{eqnarray}
with the screened interaction $W(12)$ in the 
random-phase approximation (RPA) and the free electron-hole pair propagator
\begin{equation}
K^{\sigma\sigma^{\prime}}(12,34) = \textrm{i} G^{\sigma}(13) G^{\sigma^{\prime}}(42).\label{K}
\end{equation}
Equation (\ref{eq::TmatrixBSE}) defines the $T$ matrix to contain the ladder
diagrams to all orders starting from the third order in $W$.

\section{Goldstone contribution}\label{sec:Goldstone}

As already mentioned, the $T$ matrix is closely related to the transverse
magnetic response function $R^{+-}(12)=\delta m^+(1) / \delta B^+(2)$, which
gives the linear response of the spin density with respect to an externally 
applied circularly polarized magnetic $B$ field. For simplicity, we omit 
``$+$'' and ``$-$'' denoting the circular polarizations in the following.
We now generalize the spin density to the spin density matrix $m(12)$
and the magnetic field to a non-local field $B(34)$. Then, the four-point
response function $R(12,34)=\delta m(12)/\delta B(34)$ can be written as the
solution of a BSE (see, e.g., Ref.~\onlinecite{friedrich18}
\footnote{$R=-2R^{(4)}$ in the notation of Ref.~\onlinecite{friedrich18}.}). 
In matrix notation, the solution reads
$R=-2(\mathds{1}-KW)^{-1}K$ with the free electron-hole pair propagator
of Eq.~(\ref{K}), which has poles at the energies of the unrenormalized Stoner excitations.
The inverse $(\mathds{1}-KW)^{-1}$ renormalizes these Stoner excitations, but it also gives rise to the
collective spin-wave modes. Without it, one would simply obtain $-2K$, the
transverse magnetic response function of the non-interacting mean-field system.

An important special case is the limit $\mathbf{k}\rightarrow\mathbf{0}$.
In this limit (and in the absence of spin-orbit coupling), a static $B$ field that is orthogonal to the spin
polarization is able to rotate the spins collectively towards the field direction
even if $B$ is infinitesimally small. This leads to a spin-wave delta peak in
$R$ at $\mathbf{k}=\mathbf{0}$ and $\omega=0$. For finite $\mathbf{k}$, the
spin-wave dispersion can be shown \cite{moriya85} to be quadratic in $k$, which, in a more general
sense, corresponds to the Goldstone mode expected to arise in the case of a
spontaneously broken symmetry \cite{nambu60,goldstone61} such as ferromagnetism. 
The delta peak at $\omega=0$ results from the infinite summation of ladder
diagrams and will thus also appear in the $T$ matrix, where it is likely to pose a numerical problem 
in the evaluation of the self-energy. Therefore, we discuss its
contribution to Eq.~(\ref{eq::ElMagSelfEnergy}) in the following. 

As a first
step, we have to characterize the Goldstone delta peak in a more detailed
way. As already mentioned, an externally applied infinitesimal transversal $B$ field
can rotate the electron spins macroscopically. In order for this rotation to
be rigid (collective rotation of the spins), the linear transversal change of the ground-state spin density matrix $m(12)$ must be
proportional to $m(12)$ itself, $\delta m\propto m$ (intercept theorem).
Thus, integrating the response function $R$ with the infinitesimal $B$ field should give a finite
response proportional to $m$. Conversely, the
inverse of $R$ should have a vanishing eigenvalue with the eigenvector
$m$, i.e., $(\mathds{1}-KW)m=0$. We now sketch a proof that this is fulfilled
if the Coulomb-hole screened-exchange (COHSEX) self-energy is taken as the starting point. For brevity, we use a simplified
notation. The static COHSEX self-energy
$\Sigma^\sigma(12)=-W(12)n^\sigma(12)+\Sigma_\mathrm{c}(1)\delta(12)$ with the density
matrix $n^\sigma(12)$
consists of the screened exchange (Hartree-Fock exchange potential with a statically
screened Coulomb potential) and a static Coulomb-hole self-energy, which has the
form of a local potential.
When separating off the spin-independent part
[$(\Sigma^{\uparrow}+\Sigma^{\downarrow})/2$], the remaining spin-dependent part can
formally be interpreted as arising from a nonlocal magnetic field
$B(12)=-W(12)m(12)/2$, which gives rise to the spin polarization in the COHSEX mean-field solution.
Now we use the simple fact that rigidly rotating this $B$ field 
will rotate the magnetization in the same way, implying the
proportionalities $\delta B\propto B$ and $\delta m\propto m$ with 
$\delta B/B=\delta m/m$.
With the magnetic response function $-2K$ of the mean-field system, corresponding to
the formula for $R$ without the inverse, we can formulate this
magnetic response as $-2K\delta B=\delta m$ (in matrix notation), and, with the above 
relationship for $\delta B$ and $\delta m$, we thus have $-2KB=m$.
Then, inserting $B=-Wm/2$ finally yields $KWm=m$. This proves that $m$ is an
eigenvector of $\mathds{1}-KW$ with vanishing eigenvalue, a claim that we have
previously \cite{mueller16} substantiated with arguments of theoretical
consistency. There is another more detailed proof, which is
presented elsewhere \cite{friedrich18}. 

To continue, we have to define a basis in which the quantities are to be
expanded. We will later use a Wannier product basis, but, for the present
purpose, it is easier to consider products of single-particle wave functions 
$\{
\varphi_{\mathbf{k}m}^{\uparrow}(1)
\varphi_{\mathbf{k}'m'}^{\downarrow*}(2)\}$
as a basis, instead. The Goldstone condition $KWm=m$ gives $m$ as an eigenvector 
in this basis. With $m(12)=n^\uparrow(12)-n^\downarrow(12)$ and
\begin{equation}
n^\sigma(12)=\frac{1}{N}\sum_\mathbf{k}\sum_m f^\sigma_{\mathbf{k}m}
\varphi^\sigma_{\mathbf{k}m}(1)\varphi^{\sigma*}_{\mathbf{k}m}(2)
\end{equation}
with the number $N$ of $\mathbf{k}$ points and the occupation numbers
$f^\sigma_{\mathbf{k}m}$,
we obtain after expansion of $\varphi_{\mathbf{k}m}^\sigma(1)$ in terms of
$\{\varphi^{-\sigma}_{\mathbf{k}m'}(1)\}$
\begin{equation}
m(12)=\frac{1}{N}\sum_{\mathbf{k}}\sum_{m,m'}(f_{\mathbf{k}m'}^{\uparrow}-f_{\mathbf{k}m}^{\downarrow})
\langle\varphi_{\mathbf{k}m}^{\downarrow}|\varphi_{\mathbf{k}m'}^{\uparrow}\rangle
\varphi_{\mathbf{k}m}^{\downarrow}(1)\varphi_{\mathbf{k}m'}^{\uparrow*}(2)\,.\label{m-expand}
\end{equation}
According to this expression, we can immediately conclude
(because of the factor
$f_{\mathbf{k}m}^{\uparrow}-f_{\mathbf{k}m'}^{\downarrow}$)
that the nonzero eigenvector elements are those where one of the states
$(\mathbf{k}m\uparrow)$ or ($\mathbf{k}m'\downarrow$) is occupied
and the other is unoccupied.
If the spin-up and spin-down states are not too different from each
other, we can also see that the combination $(\mathbf{k}m\uparrow)$
occupied and ($\mathbf{k}m'\downarrow$) unoccupied gives a sizable
contribution, while the other term is negligible (because of the small overlap).

These considerations are helpful in the discussion of the Goldstone-mode
contribution to the $GT$ self-energy
\begin{equation}
\label{convol}
\Sigma^{\sigma}(\omega)=-\frac{\mathrm{i}}{2\pi}\int_{-\infty}^\infty
T^{\sigma\sigma'}(\omega')G^{\sigma'}(\omega-\omega')\mathrm{d}\omega'\,,
\end{equation}
here written in simplified notation and Fourier space.
The frequency
integral can be evaluated along a closed contour from $-\infty$ to $\infty$
and back to $-\infty$ along an infinite semiarc in the upper or lower
complex half-plane. The latter does not contribute because both the $T$ matrix
and the Green function behave asymptotically as $\omega^{-1}$.
When writing $G$ and $T$ as sums over poles [e.g., see
Eq.~(\ref{Lehmann0})], the contribution of any pair of poles can be evaluated
with the residue theorem, and it vanishes unless the poles
are both below or both above the real frequency axis
(note the negative sign of $\omega'$ in the Green function).

Let us consider the spin-down case first. The $T^{\downarrow\uparrow}$ matrix has
the Goldstone delta peak infinitesimally below the real frequency axis, so
its contribution is proportional to
$1/(\omega-\epsilon_\mathrm{M}+\mathrm{i}\eta)$
with $\epsilon_\mathrm{M}=0$. 
Then, in the frequency convolution of Eq.~(\ref{convol}), 
only the poles of the Green function that are also below the real
axis contribute, which, according to Eq.~(\ref{Lehmann0}), are those of the unoccupied states
($\mathbf{k}m'\uparrow$), resulting in self-energy poles at
$\epsilon_{\mathbf{k}m'}^{\uparrow}+\epsilon_\mathrm{M}=\epsilon_{\mathbf{k}m'}^{\uparrow}$.
According to Eq.~(\ref{m-expand}) and the discussion thereafter,
the unoccupied states ($\mathbf{k}m'\uparrow$) must be combined with
occupied states ($\mathbf{k}m\downarrow$) to form the eigenvector of the $T$
matrix that is related to the Goldstone peak. The corresponding
eigenvector elements are exactly those that have been discussed above to be very small.
Furthermore, because of the combination of states, only the self-energy matrix elements
that couple occupied states can get a non-zero contribution from this
eigenvector, for example, the diagonal element
$\Sigma_{\mathbf{k}m}^\downarrow(\omega)=\langle\varphi_{\mathbf{k}m}^{\downarrow}|\Sigma^{\downarrow}(\omega)|\varphi_{\mathbf{k}m}^{\downarrow}\rangle$.
The contribution would consist of poles located at frequencies larger than
the Fermi energy,
thus 
in an energy region that is less
relevant for the renormalization of the 
occupied states ($\mathbf{k}m\downarrow$).
Analogous arguments apply to the spin-up case:
The $T^{\uparrow\downarrow}$ matrix has its Goldstone delta peak
infinitesimally above the real-frequency axis. So, it is the
Green-function poles related to the occupied states
($\mathbf{k}m\downarrow$) that now give rise to self-energy poles at
$\epsilon_{\mathbf{k}m}^{\downarrow}$,
the other poles do not contribute. These states have to be combined with the
unoccupied states ($\mathbf{k}m'\uparrow$)---the same combination as before---to yield a contribution to the
diagonal element
$\Sigma_{\mathbf{k}m'}^\uparrow(\omega)=\langle\varphi_{\mathbf{k}m'}^{\uparrow}|\Sigma^{\uparrow}(\omega)|\varphi_{\mathbf{k}m'}^{\uparrow}\rangle$
with poles at frequencies below the Fermi energy. Again, these poles do not
play an important role for the renormalization of the unoccupied states
($\mathbf{k}m'\uparrow$).

The Goldstone contribution to the $GT$ self-energy
is thus expected to be small.
Of course, the spin-wave peaks at finite $\mathbf{k}$ and $\omega$ can
contribute sizably to the self-energy, and we will see that these collective
excitations can give rise to anomalies in the electronic band dispersions.



\section{Implementation} \label{sec::Implementation}
We have implemented the $GT$ self-energy into the $GW$ code
\textsc{spex}~\cite{friedrich10}.
The self-consistent DFT calculations are carried out with the \textsc{fleur} code~\cite{fleur}.
Both codes rely on the all-electron full-potential linearized
augmented-plane-wave method, which provides an accurate basis set for the 
representation of both the itinerant $s$ and $p$ states as well as the
localized $d$ states.
Solving the BSE [Eq.~(\ref{eq::TmatrixBSE})] leads to an
infinite series of ladder diagrams. This series appears also in the diagrammatic
expansion of the transverse magnetic response function (then starting from
the first order in $W$), for which we
reported an implementation in the Refs.~\onlinecite{sasioglu10}-\onlinecite{mueller16}.
The implementation exploits the fact that the Hamiltonian is not explicitly
time dependent. In addition, the RPA screened interaction $W(\mathbf{r},\mathbf{r}';\omega)$ is approximated
by its static limit $W(\mathbf{r},\mathbf{r}')=W(\mathbf{r},\mathbf{r}';0)$. 
As a result, the self-energy [Eq.~(\ref{eq::ElMagSelfEnergy})] involves only 
a single frequency integration. Furthermore, we restrict ourselves
to calculating only the diagonal elements 
$\Sigma^\sigma_{\mathbf{q}m}=\langle \varphi_{\mathbf{q}m}^{\sigma}|
\Sigma^{\sigma}(\omega)|\varphi_{\mathbf{q}m}^{\sigma}\rangle$
of the self-energy
\begin{eqnarray}
\Sigma^\sigma_{\mathbf{q}m}(\omega)=
-\frac{\textrm{i}}{2\pi} 
\int \textrm{d}^3 r_1\, \textrm{d}^3 r_2\, \textrm{d}^3 r_3\, \textrm{d}^3
r_4 \,\,
\varphi_{\mathbf{q}m}^{\sigma *}(\mathbf{r}_1)
\varphi_{\mathbf{q}m}^{\sigma}(\mathbf{r}_{3})
\nonumber\\
\times\int_{-\infty}^{\infty} 
T^{\sigma\sigma^{\prime}}(\mathbf{r}_{1},\mathbf{r}_{2};\mathbf{r}_{3},\mathbf{r}_{4};\omega^{\prime})
G^{\sigma^{\prime}}(\mathbf{r}_{2},\mathbf{r}_{4};\omega-\omega^{\prime})
\textrm{d}\omega^{\prime}\,\,
\label{GTimp}
\end{eqnarray}
with the momentum $\mathbf{q}$, the band index $m$, and the spin $\sigma$ of
the Bloch state $\varphi_{\mathbf{q}m}^{\sigma}(\mathbf{r})$.
The short-range behavior of the screened interaction $W$ in metals allows an
on-site approximation to be employed
\cite{sasioglu10,friedrich14,mueller16,friedrich18}: The electron and hole are
assumed to occupy the same lattice site when they interact with each other.
To realize a lattice-site resolution, we formulate the theory with the help of 
Wannier functions
\begin{equation}
w_{\mathbf{R} n}^{\sigma} (\mathbf{r}) = \frac{1}{N}\sum_{\mathbf{k}}
e^{-\textrm{i}\mathbf{k}\mathbf{R}} \sum_{m} U_{\mathbf{k}m,n}^{\sigma} \varphi_{\mathbf{k}m}^{\sigma}(\mathbf{r}),
\end{equation}
where $\mathbf{R}$ is the lattice site, $n$ is an orbital index,
and $N$ is the number of 
$\mathbf{k}$ points. The complex expansion coefficients
$U_{\mathbf{k}m,n}^{\sigma}$ of the $n$th Wannier orbital 
are given by projection of the eigenstates onto a muffin-tin basis function with
suitable orbital character and a subsequent L\"owdin orthonormalization
\cite{freimuth08}.
The Wannier representation of the $T$ matrix reads
\begin{eqnarray}
T^{\sigma\sigma^{\prime}}(\mathbf{r}_{1},\mathbf{r}_{2};\mathbf{r}_{3},\mathbf{r}_{4};\omega) = \sum_{\mathbf{R},\mathbf{R}^{\prime}} 
w_{\mathbf{R} n_{1}}^{\sigma} (\mathbf{r}_{1}) w_{\mathbf{R} n_{2}}^{\sigma^{\prime} *} (\mathbf{r}_{2}) \nonumber \\
\times
T^{\sigma\sigma^{\prime}}_{\mathbf{R}n_{1}\mathbf{R}n_{2},\mathbf{R}^{\prime}n_{3}\mathbf{R}^{\prime}n_{4}}(\omega)
w_{\mathbf{R}^{\prime} n_{3}}^{\sigma *} (\mathbf{r}_{3})
w_{\mathbf{R}^{\prime} n_{4}}^{\sigma^{\prime} } (\mathbf{r}_{4}).
\end{eqnarray}
Due to the lattice periodicity, the coefficients
$T^{\sigma\sigma^{\prime}}_{\mathbf{R}n_{1}\mathbf{R}n_{2};\mathbf{R}^{\prime}n_{3}\mathbf{R}^{\prime}n_{4}}(\omega)$
depend only on the difference vector $\Delta\mathbf{R}=\mathbf{R}-\mathbf{R'}$. A lattice
Fourier transformation then yields
\begin{equation}
T^{\sigma\sigma'}_{n_1n_2,n_3n_4}(\mathbf{k},\omega)=\sum_{\Delta\mathbf{R}}T^{\sigma\sigma'}_{n_1n_2,n_3n_4;\Delta\mathbf{R}}(\omega)e^{-\mathrm{i}\mathbf{q}\Delta\mathbf{R}}.
\end{equation}
Employing the Lehmann representation of the Green function
\begin{equation}\label{Lehmann0}
G^{\sigma}(\mathbf{r},\mathbf{r}^{\prime};\omega) = \frac{1}{N} \sum_{\mathbf{k}}\sum_{m}
\frac{\varphi_{\mathbf{k}m}^{\sigma}(\mathbf{r}) \varphi_{\mathbf{k}m}^{\sigma *}(\mathbf{r}^{\prime})}
{\omega - \epsilon_{\mathbf{k}m}^{\sigma} + \textrm{i}\eta\,\textrm{sgn}(\epsilon_{\mathbf{k}m}^{\sigma}-\epsilon_{\textrm{F}})}
\end{equation}
with the energies $\epsilon_{\mathbf{k}m}^{\sigma}$ of the corresponding Bloch
states, the Fermi energy $\epsilon_{\textrm{F}}$, and a positive
infinitesimal $\eta$, one obtains for the diagonal element of the $GT$ self-energy
\begin{eqnarray}\label{eq::DiagonalSelfEnergyRealFrequency}
\Sigma_{\mathbf{q}m}^{\sigma}(\omega) &=& -\frac{\textrm{i}}{2\pi}
\int_{-\infty}^{\infty}\textrm{d}\omega^{\prime}\;\sum_{\mathbf{k}} \sum_{n_{1},n_{2},n_{3},n_{4}} T_{n_{1}n_{2},n_{3}n_{4}}^{\sigma\sigma^{\prime}}(\mathbf{k},\omega^{\prime}) \nonumber \\
&\times& \sum_{m'} \frac{ 
U_{\mathbf{q}m,n_{1}}^{\sigma} U_{\mathbf{q}-\mathbf{k}m',n_{2}}^{\sigma^{\prime} *}
U_{\mathbf{q}m,n_{3}}^{\sigma *} U_{\mathbf{q}-\mathbf{k}m',n_{4}}^{\sigma^{\prime}}
}
{
\omega-\omega^{\prime}-\epsilon_{\mathbf{q}-\mathbf{k}m'}^{\sigma^{\prime}} + \textrm{i}\eta\,\textrm{sgn}(\epsilon_{\mathbf{q}-\mathbf{k}m'}^{\sigma^{\prime}}-\epsilon_{\textrm{F}})
},
\end{eqnarray}
where the summation over $m'$, in principle, runs over the infinitely
many single-particle
eigenstates, but in the present case the summation is effectively restricted to the ones which have been used for the construction of the Wannier basis. 

The $T$ matrix exhibits poles at the spin
excitation energies along the real-frequency axis. The collective spin-wave excitations produce
particularly strong poles at low energies, which complicates a straightforward
frequency integration.
The integration along the real-frequency 
axis can be avoided by employing the method of analytic continuation
\cite{rojas95,rieger99}, which is one of the two methods implemented.
To this end, Eq.~(\ref{eq::DiagonalSelfEnergyRealFrequency}) is
analytically continued to the imaginary axis and evaluated there for a set
of imaginary frequencies forming a mesh along this axis.
This is achieved by replacing $\omega \to \textrm{i}\omega$ and
$\omega' \to \textrm{i}\omega'$, formally changing the prefactor
$-\textrm{i}/(2\pi)$ to $1/(2\pi)$.
In this way, one avoids the real-frequency axis where the quantities show
strong variations. Along the imaginary axis, the functions are much
smoother, which enables an accurate sampling and interpolation of the
functions with relatively coarse frequency meshes, also simplifying the
frequency convolution of Eq.~(\ref{eq::DiagonalSelfEnergyRealFrequency}).
At the end of the calculation, the self-energy has to be analytically
continued back to the real-frequency axis.
We employ Pad\'{e} approximants for the $T$ matrix (Thiele's
reciprocal difference method \cite{baker10}), which allows an analytic
frequency convolution with $G$, and also for the self-energy, yielding the
self-energy on the whole complex frequency plane. 
In this approach, the self-energy is effectively represented by a sum over poles 
with analytically determined complex positions and weights. However, it is well
known that this extrapolation can 
lead to spurious features in the self-energy if one of 
the effective poles happens to lie close to the real frequency axis. 

\cf{

Therefore, we have also implemented the contour deformation technique to
evaluate the self-energy. This method is more accurate than analytic
continuation but also requires more computation time. Details of both
implementations will be presented elsewhere \cite{friedrich19} together 
with a new tetrahedron method, with which the two methods yield nearly identical results. There, we
also discuss a third method that could be described as a hybrid of the two.
The contour deformation technique evaluates the frequency integral of
Eq.~(\ref{eq::DiagonalSelfEnergyRealFrequency}) explicitly, yielding the
self-energy directly for real frequencies. The integration path, however,
does not run along the real frequency
axis from $-\infty$ to $\infty$ but along a deformed integration contour:
from $-\mathrm{i}\infty$ to $\mathrm{i}\infty$ and, at halfway, taking a
\emph{rectangular detour} at $\omega=0$ to capture all residues of the Green
function. (The integrations from $-\infty$ to $-\mathrm{i}\infty$ and from
$\mathrm{i}\infty$ to $\infty$ do not contribute.) The resulting self-energy
can be written as a sum of two terms
\begin{eqnarray}
\label{contdef}
\lefteqn{\langle\varphi^\sigma_{\mathbf{q}m}\left|\Sigma^\sigma(\omega)\right|\varphi^\sigma_{\mathbf{q}m}\rangle}
\\
\nonumber
&=& 
\frac{1}{N} \sum_\mathbf{k} \left [ \sum_{m'} \frac{1}{2\pi}
\int_{-\infty}^{\infty}d\omega'\frac{T^{\sigma\sigma'}_{\mathbf{q}mm'}(\mathbf{k},-\mathrm{i}\omega')}{\omega+\mathrm{i}\omega'-\epsilon^{\sigma'}_{\mathbf{q-k}m'}}
\right . \\
&+& \left . \left\{ \begin{array}{rr}
 {\displaystyle \sum_{\substack{m' \\ \omega<\epsilon^{\sigma'}_{\mathbf{q-k}m'}<\epsilon_\mathrm{F}}}}
 T^{\sigma\sigma'}_{\mathbf{q}mm'}(\mathbf{k},\omega-\epsilon^{\sigma'}_{\mathbf{q-k}m'}) & \textrm{ for }\omega<\epsilon_{\mathrm{F}}\\
-{\displaystyle \sum_{\substack{m' \\ \epsilon_\mathrm{F}<\epsilon^{\sigma'}_{\mathbf{q-k}m'}<\omega}}}
 T^{\sigma\sigma'}_{\mathbf{q}mm'}(\mathbf{k},\omega-\epsilon^{\sigma'}_{\mathbf{q-k}m'}) & \textrm{ for }\omega>\epsilon_{\mathrm{F}}\end{array}\right.\right ] \,,
\nonumber
\end{eqnarray}
where $T^{\sigma\sigma'}_{\mathbf{q}mm'}(\mathbf{k},\omega)$ denotes the sum of the products $TUUUU$ 
over $n_1$, $n_2$, $n_3$, and $n_4$. The $\mathbf{k}$ summation should be
understood as a $\mathbf{k}$ integration. It is important to perform this
integration accurately. We have devised a new tetrahedron method for this
purpose \cite{friedrich19}. All results presented in this paper have been
calculated with the contour deformation technique using Eq.~(\ref{contdef}).

}



The evaluation of Eq.~(\ref{contdef}) can be
expensive. Therefore, we exploit a few symmetries to accelerate the
computation. The first is a symmetry in frequency space which the $T$ matrix 
inherits from the propagator $K$ \cite{friedrich14},
$T_{n_{1}n_{2},n_{3}n_{4}}^{\sigma\sigma^{\prime}}(\mathbf{k},-\mathrm{i}\omega)
=T_{n_{3}n_{4},n_{1}n_{2}}^{\sigma\sigma^{\prime}*}(\mathbf{k},\mathrm{i}\omega)$.
This leads to
$T_{\mathbf{q}mm'}^{\sigma\sigma'}(\mathbf{k},-\mathrm{i}\omega)=
T_{\mathbf{q}mm'}^{\sigma\sigma'*}(\mathbf{k},\mathrm{i}\omega)$.
Hence, we can restrict the frequency
mesh to the positive imaginary axis and utilize this symmetry for the
frequency integration from $-\mathrm{i}\infty$ to $\mathrm{i}\infty$. \cf{Furthermore, it follows
from this symmetry that the first term of Eq.~(\ref{contdef}) is 
real-valued. The second term is complex in general, which will become important in
Sec.~\ref{sec:causality}.}
We also exploit spatial symmetry in the evaluation of the $\mathbf{k}$ summations.
Equation (\ref{eq::DiagonalSelfEnergyRealFrequency}) contains a
sum over all $\mathbf{k}$ vectors. We restrict this summation to the
extended irreducible Brillouin zone (corresponding to the irreducible zone
that is created by the subset of symmetry operations that leave the $\mathbf{q}$
vector invariant \cite{friedrich10}, the so-called little group).
The contribution of the symmetry-equivalent $\mathbf{k}$ points is obtained
by a symmetrization procedure employing the symmetry transformation matrices
of the Wannier set \cite{friedrich14}. In the same way, we can accelerate the computation of
the two-particle propagator [Eq.~(\ref{K})], which involves a
$\mathbf{k}$ summation, as well \cite{friedrich14}.

As already mentioned above, we presently refrain from combining
Eq.~(\ref{eq::ElMagSelfEnergy}) with the fully dynamical $GW$ self-energy.
Such a study is deferred to a later work. Instead, 
we combine it with the
LSDA solution whose exchange splitting has been corrected in such a way
that the Goldstone condition (the spin-wave energy must vanish in the limit 
$\mathbf{k}\rightarrow 0$) is fulfilled. To be more precise, we introduce a
parameter $\Delta_\mathrm{x}$, with which the spin-up and spin-down states are
shifted with respect to each other, i.e., 
$\epsilon^{\uparrow/\downarrow}_{\mathbf{k}m}\rightarrow
\epsilon^{\uparrow/\downarrow}_{\mathbf{k}m} \pm \Delta_\mathrm{x}/2$. (A positive
$\Delta_\mathrm{x}$ decreases the exchange splitting.) The parameter $\Delta_\mathrm{x}$
is varied until the spin-wave excitation energy vanishes in the long-wavelength limit.
The so-corrected single-particle system has been shown~\cite{mueller16} to be similar 
to a self-consistent solution obtained with the COHSEX
self-energy~\cite{hedin65}, the static limit of $GW$. The corrected LSDA
Green function can thus be understood as an approximation to the COHSEX
Green function, which is then further renormalized with the $GT$ self-energy.


We introduce another parameter to achieve partial self-consistency in the
Green function.
The application of the $GT$ self-energy does not only introduce lifetime
broadening effects but also leads to an energetic shift of the electronic
bands. In general, this entails that the Fermi energy has to be readjusted
to ensure particle number conservation. Such a readjustment,
however, has some undesired consequences. First, the spectral function
vanishes at the old Fermi energy but not at the new one. Second and more
seriously, spectral features, e.g., kinks in the quasiparticle bands, might
end up at a wrong binding energy, in particular, if their energetic position
is of the same order of magnitude as the shift of the Fermi energy. In an
extreme case, the binding energy might even change sign.
Clearly, in a fully self-consistent $GT$ calculation, the Fermi energy would
converge eventually, and these problems would be solved. However, presently we cannot
afford such a calculation. 
Therefore, we employ a correction that was originally proposed by Hedin
\cite{hedin65} and discussed in detail in Refs.~\onlinecite{schindlmayr97}
and \onlinecite{pollehn98}. In a one-shot calculation, there is a mismatch between
the chemical potential of the non-interacting system, inherited by the self-energy, 
and the chemical potential of the renormalized Green function.
In order to align the two potentials, one introduces an energy shift
$\Delta_v$ in
$\Sigma(\omega) \rightarrow \Sigma(\omega-\Delta_v)$, where $\Delta_v$ is chosen
in such a way that the Fermi energy remains unchanged by the self-energy
renormalization, giving the spectral function as
\begin{equation}
S^\sigma(\mathbf{k},\omega)=\frac{\mathrm{sgn}(\epsilon_\mathrm{F}-\omega)}{\pi}
\sum_m\textrm{Im}\{\omega-\epsilon^\sigma_{\mathbf{k}m}-\Sigma^\sigma_{\mathbf{k}m}(\omega-\Delta_v)\}^{-1}.
\label{Dyson0}
\end{equation}
Since the corrected energies approximate COHSEX quasiparticle energies
and COHSEX does not share any diagrams with $GT$,
no subtraction of the expectation value of the exchange-correlation
potential as in one-shot $GW$ calculations is needed in this case.
\cf{The spectral function in Eq.~(\ref{Dyson0}) is defined as the trace over
the imaginary part of the renormalized Green function, the solution 
of the Dyson equation with the $GT$ self-energy. The sign is chosen
such that the spectral function is non-negative for all $\omega$ and
$\mathbf{k}$ (see Sec.~\ref{sec:causality}).}

The correction is tantamount to shifting the 
exchange-correlation potential by the same amount $v_\mathrm{xc}^\sigma \rightarrow
v_\mathrm{xc}^\sigma+\Delta_v$. In this way, the non-interacting reference system 
already contains information about the renormalized
system through the shift $\Delta_v$, which can be regarded as some level of
self-consistency.
(Since the Fermi energy depends on the quasiparticle energies in the whole
irreducible wedge of the Brillouin zone, the shift has to be calculated
self-consistently until the Fermi energy and $\Delta_v$ converge.)
This approach solves the aforementioned two problems in a natural way. 

The modification of $v^\sigma_\mathrm{xc}$ changes the single-particle energies
accordingly so that, in summary, the latter are adjusted by 
$\epsilon^{\uparrow/\downarrow}_{\mathbf{k}m}\rightarrow
\epsilon^{\uparrow/\downarrow}_{\mathbf{k}m} + \Delta_v \pm
\Delta_\mathrm{x}/2$, which shifts the self-energy argument in the opposite
direction, effectively undoing the shift from before,
$\Sigma(\omega-\Delta_v)\rightarrow\Sigma(\omega)$.
The $\Delta_v$ correction does not
affect the Goldstone condition, which has been used to fix
$\Delta_\mathrm{x}$, since only the differences of the single-particle
states (of opposite spins) enter the evaluation of the spin excitations
energies, and not their absolute values. 
We note that the two parameters, $\Delta_\mathrm{x}$ and $\Delta_v$, are 
determined from exact physical constraints, so they do not imply a
deviation from the ``ab initio-ness'' of the method.

With the aligned chemical potential, Eq.~(\ref{Dyson0}) can alternatively
be written as
\begin{equation}
S^\sigma(\mathbf{k},\omega)=\frac{\mathrm{sgn}(\epsilon_\mathrm{F}-\omega)}{\pi}
\sum_m\textrm{Im}\{\omega-\epsilon^\sigma_{\mathbf{k}m}-\Sigma^\sigma_{\mathbf{k}m}(\omega)+\Delta_v\}^{-1},
\label{Dyson}
\end{equation}
where the single-particle energies $\epsilon^\sigma_{\mathbf{k}m}$ are
assumed to contain the two parameters $\Delta_\mathrm{x}$ and $\Delta_v$ as
described above.
(We note that the parameter $\Delta_v$ in Eq.~(\ref{Dyson}) seems to have no
effect because the term $-\Delta_v$ de facto cancels the $\Delta_v$ in
$\epsilon^\sigma_{\mathbf{k}m}$, but it should be remembered that $\Delta_v$
shifts the Fermi energy, too. We also note that it might be surprising
that a spin-independent shift of the exchange-correlation potential and,
thus, of the Kohn-Sham eigenvalues should have an
effect at all, but one has to remember that quasiparticle energies
correspond to total-energy differences and are therefore defined absolutely.)

\cf{

\section{Violation of causality}\label{sec:causality}

In this section, we present a detailed discussion about the violation of causality mentioned in
the introduction. To this end, we have to resort to a general
second-quantization formulation of the Green function and related response quantities.
The Bloch $\mathbf{k}$ crystal momentum is suppressed for simplicity.
The time-ordered Green function
\begin{equation}\label{G}
G^\sigma(12)=-\mathrm{i}
\langle\Psi_0|{\cal T}[\hat{\psi}^\sigma(1)\hat{\psi}^{\sigma\dagger}(2)]|\Psi_0\rangle
\end{equation}
comprises both the electron and the hole propagator by virtue of the
time-ordering operator $\cal T$: If $t_1>t_2$, the Heisenberg creation operator
$\hat{\psi}^{\sigma \dagger}(2)$ acts first on the many-body ground state
$|\Psi_0\rangle$ creating an electron at 2, which is later
annihilated at 1. If otherwise $t_1<t_2$, then the operator $\cal T$ switches
the order of the field operators and lets $\hat{\psi}^\sigma(1)$ act first, thus creating a hole at 1, which is later
filled by an electron at 2. So, the Green function describes the propagation
of an electron or a hole depending on the time order of the arguments $t_1$ and $t_2$.
The Fourier transformation of Eq.~(\ref{G}) ($t_1-t_2\rightarrow\omega$) yields the Lehmann representation 
\footnote{To sketch the derivation, which can be found in many textbooks,
e.g., Ref.~\onlinecite{mahan00}, we write Eq.~(\ref{G}) as the sum of two
terms corresponding to the cases $t_1>t_2$ and $t_1<t_2$ with the help of the Heaviside function
$\theta(t_1-t_2)=1$ for $t_1>t_2$ and 0 otherwise. \cf{(It should be noted
that, in the fermionic case, 
the operator $\cal{T}$ is defined to introduce a factor $-1$ when it switches
the field operators.)}
Then, inserting the closure
relation $\sum_m
|\Psi_m^{\sigma\pm}\rangle\langle\Psi_m^{\sigma\pm}|=\hat{1}$ between the
two field operators in each term,
using
$\hat{\psi}^\sigma(\mathbf{r}t)=e^{\mathrm{i}\hat{H}t}\hat{\psi}^\sigma_\mathrm{S}(\mathbf{r})e^{-\mathrm{i}\hat{H}t}$
and
$\int\theta(t)e^{\mathrm{i}\omega t}dt=\mathrm{i}/(\omega+\mathrm{i}\eta)$
yields Eq.~(\ref{Lehmann}).}
\begin{eqnarray}
\nonumber
G^{\sigma}(\mathbf{r}_1,\mathbf{r}_2;\omega) &=&
\sum_m
\frac{
\langle\Psi_0|\hat{\psi}_\mathrm{S}^\sigma(\mathbf{r}_1)|\Psi_{m}^{\sigma+}\rangle
\langle\Psi_{m}^{\sigma+}|\hat{\psi}_\mathrm{S}^{\sigma\dagger}(\mathbf{r}_2)|\Psi_0\rangle
}
{\omega - ( E_{m}^{\sigma+} - E_0 ) + \textrm{i}\eta} \\
&+&\sum_m
\frac{
\langle\Psi_{m}^{\sigma-}|\hat{\psi}_\mathrm{S}^\sigma(\mathbf{r}_1)|\Psi_0\rangle
\langle\Psi_0|\hat{\psi}_\mathrm{S}^{\sigma\dagger}(\mathbf{r}_2)|\Psi_{m}^{\sigma-}\rangle
}
{\omega - ( E_0 - E_{m}^{\sigma-} ) - \textrm{i}\eta}
\,,
\label{Lehmann}
\end{eqnarray}
where $\eta$ is a positive infinitesimal and
$\{|\Psi_m^{\sigma\pm}\rangle,E_m^{\sigma\pm}\}$ is a complete set of
energy eigensolutions of the many-body system with one spin-$\sigma$ electron more ($+$) or less ($-$)
than $|\Psi_0\rangle$.

We now consider a diagonal element 
$G^\sigma_m(\omega)\coloneqq\langle\varphi_m^\sigma|G^\sigma(\omega)|\varphi_m^\sigma\rangle$ 
with $\varphi_m^\sigma(\mathbf{r})$ an eigenstate of the non-interacting
reference system.
Since the Schr\"odinger field operators 
$\hat{\psi}^\sigma_\mathrm{S}(\mathbf{r})$ and
$\hat{\psi}^{\sigma\dagger}_\mathrm{S}(\mathbf{r})$ are adjoints of each other, 
the numerators, when projected onto $\varphi_m^\sigma(\mathbf{r})$, are real
and non-negative. Furthermore, 
$E_0 - E_{m}^{\sigma-} \le \mu \le E_{m}^{\sigma+} - E_0$ with the chemical
potential $\mu$. Then, it follows from $(\omega\pm\mathrm{i}\eta)^{-1}={\cal
P}(1/\omega)\mp\mathrm{i}\pi\delta(\omega)$ that the imaginary part of
$G^\sigma_m(\omega)$
must have a positive (negative) sign for $\omega<\mu$ ($\omega>\mu$). 
As $\eta$ enters Eq.~(\ref{Lehmann}) as a consequence of the time-ordering operator, 
we can trace the condition on the sign of $\mathrm{Im}G^\sigma_m(\omega)$ back to the correct
time order of the field operators and thus to the causality. 

The mathematical formulas derived above bear similarity with those
encountered in the treatment of the \emph{damped driven harmonic oscillator}. In
this context, the energy differences would correspond to the resonance
frequency of the free oscillator and $\eta$ could be interpreted as a damping
coefficient. The latter appears linearly in the imaginary part of the
free oscillator's frequency.
So, a wrong sign in the imaginary part presupposes a wrong sign
in $\eta$, which would thus have the effect of amplifying the oscillation rather
than damping it. This solution could be described as one that goes backwards in
time, thus violating causality.

Through the Dyson equation, the imaginary part of
$G^\sigma_m(\omega)$ can be related to the
imaginary part of the self-energy
\begin{equation}
\label{GreenSigma}
\mathrm{Im}G^\sigma_m(\omega)=\frac{\mathrm{Im}\Sigma^\sigma_m(\omega)}{[\omega-\epsilon_m^\sigma-\mathrm{Re}\Sigma^\sigma_m(\omega)]^2+[\mathrm{Im}\Sigma_m^\sigma(\omega)]^2}\,,
\end{equation}
where it has been used that Eq.~(\ref{Lehmann}) is trivially valid for the
non-interacting $G$, too, in which case $E_{m}^{\sigma+} - E_0$ and 
$E_0 - E_{m}^{\sigma-}$ correspond to the single-particle energies
$\epsilon_m^\sigma$ and the numerators are just products of single-particle
wave functions
$\varphi_m^\sigma(\mathbf{r}_1)\varphi_m^{\sigma*}(\mathbf{r}_2)$.
Equation (\ref{Lehmann0}) shows such a Lehmann representation formulated
with Bloch vectors.
We have neglected the offdiagonal elements of
$\Sigma^\sigma(\omega)$ for simplicity. According to Eq.~(\ref{GreenSigma}), the sign rule for
$\mathrm{Im}G^\sigma_m(\omega)$ applies
equally to $\mathrm{Im}\Sigma^\sigma_m(\omega)$, also see Eq.~(9.34) of
\cite{FetterWalecka71}.
In the last section, it was shown [Eq.~(\ref{contdef})] that the imaginary part of 
$\Sigma^\sigma_m(\omega)$ is completely determined by 
(a $\mathbf{k}$ integration of) the imaginary part of $T$-matrix elements,
represented in terms of single-particle eigenfunctions. From
Eq.~(\ref{contdef}), it
follows that the condition on the sign of $\Sigma_m^\sigma(\omega)$ formulated
above requires the imaginary part of the $T$ matrix elements to be non-negative for
all $\omega$.

The $T$ matrix, as we defined it above, contains all ladder diagrams
starting from third order. In this sense, we can express it as
\begin{equation}
\label{T2}
T^{\sigma\sigma^\prime} = T_{\ge 2}^{\sigma\sigma^\prime} - T_2^{\sigma\sigma^\prime}\,,
\end{equation}
where $T_2$ is the bare second-order ladder diagram and $T_{\ge 2}$
includes all ladder diagrams starting from second order. 
This reformulation looks trivial (and it is), but it will help us to understand
the reason for the violation of causality. As a first step, we realize that
all ladder diagrams are flanked from either side by an interaction line
$W$, which can be factored out giving
\begin{equation}
\label{TinR}
T^{\sigma\sigma^\prime}(12,34) = W(12) [ R^{(4)\sigma\sigma^\prime}(12,34) - K^{\sigma\sigma^\prime}(12,34) ] W(34)\,.
\end{equation}
Here, $K$ is the free electron-hole propagator of Eq.~(\ref{K}) and
$R^{(4)}$ the interacting electron-hole propagator, which is related to the 
generalized four-point response function by $R=-2R^{(4)}$.
Both quantities can be expressed in terms of a spin-spin correlation
function \cite{friedrich18}
$\mathrm{i}\langle\Psi_0|{\cal T}[
\hat{\psi}^{\sigma \dagger}(2) \hat{\psi}^{\sigma'}(1)
\hat{\psi}^{\sigma'\dagger}(3) \hat{\psi}^ \sigma  (4)
]|\Psi_0\rangle$. (Note the assumption $\sigma'=-\sigma$.) For the
free (interacting) propagator $K$ ($R^{(4)}$), one would take $|\Psi_0\rangle$ to be the
energy ground state of the non-interacting reference (real interacting) system.
The spin-spin correlation function has a similar form as Eq.~(\ref{G}) and
can be Fourier-transformed analogously giving
\begin{eqnarray}
\label{RLehmann}
\lefteqn{R^{(4)\sigma\sigma'}(\mathbf{r}_1,\mathbf{r}_2,\mathbf{r}_3,\mathbf{r}_4;\omega)}\\
\nonumber
&=&
\sum_m
\frac{
\langle\Psi_{m}|
\hat{\psi}_\mathrm{S}^{\sigma\dagger}  (\mathbf{r}_2)
\hat{\psi}_\mathrm{S}^{\sigma'}        (\mathbf{r}_1)
|\Psi_0\rangle
\langle\Psi_0|
\hat{\psi}_\mathrm{S}^{\sigma'\dagger} (\mathbf{r}_3)
\hat{\psi}_\mathrm{S}^{\sigma}         (\mathbf{r}_4)
|\Psi_{m}\rangle
}
{\omega - ( E_0 - E_{m} ) - \textrm{i}\eta} \\
&-&
\sum_m
\frac{
\langle\Psi_0|
\hat{\psi}_\mathrm{S}^{\sigma\dagger}  (\mathbf{r}_2)
\hat{\psi}_\mathrm{S}^{\sigma'}        (\mathbf{r}_1)
|\Psi_{m}\rangle
\langle\Psi_{m}|
\hat{\psi}_\mathrm{S}^{\sigma'\dagger} (\mathbf{r}_3)
\hat{\psi}_\mathrm{S}^{\sigma}         (\mathbf{r}_4)
|\Psi_0\rangle
}
{\omega - ( E_{m} - E_0 ) + \textrm{i}\eta}
\,,
\nonumber
\end{eqnarray}
where $\{|\Psi_m\rangle,E_m\}$ is a complete set of
energy eigensolutions of the many-body system (with the same number of electrons as
$|\Psi_0\rangle$).

Inserting this expression and the corresponding one
for $K$ (which could be simplified again using the single-particle
eigenstates, which is, however, not necessary for the present discussion)
into Eq.~(\ref{TinR}) yields the spectral representations of the two contributions $T_{\ge 2}$ and $T_2$. The $T$
matrix elements relevant for the self-energy are obtained by multiplying
with two wave-function products,
$\varphi_{m'}^{\sigma'*}(\mathbf{r}_2)\varphi_{m'}^{\sigma'}(\mathbf{r}_4)$
and
$\varphi_m^{\sigma*}(\mathbf{r}_1)\varphi_m^\sigma(\mathbf{r}_3)$,
and integrating over all space coordinates. The first product originates
from the Green function in Eq.~(\ref{eq::ElMagSelfEnergy}) 
and the second from evaluating the diagonal element
of the self-energy $\Sigma_m^\sigma(\omega)$, also see Eq.~(\ref{GTimp}). In
analogy to the case of the Green function,
this leads to numerators in the two terms $T_{\ge 2}$ and
$T_2$ that are all real and non-negative. It is then straightforward to deduce from
Eq.~(\ref{RLehmann}) that the imaginary parts of both $T_{\ge 2}$ and
$T_2$ are non-negative in the whole frequency range.
According to Eq.~(\ref{contdef}), the imaginary part of the self-energy contributions
originating from $T_{\ge 2}$ and
$T_2$ will both have the correct
sign for $\omega>\mu$ and $\omega<\mu$ as a consequence.

Here, the decisive point is that they have the correct sign
\emph{individually}, but their difference [Eq.~(\ref{TinR})], might have a wrong sign in some frequency region and
thus violate causality.
This is not just a theoretical possibility but rather an expected behavior due to a
redistribution of spectral weight in the renormalized pair propagator $R^{(4)}$
with respect to $K$, the free pair propagator. The imaginary part of the latter is
formed by the spin-flip Stoner excitations, whereas the Bethe-Salpeter renormalization
gives rise to the formation of collective excitations (spin waves) in $R^{(4)}$ at low
energies. Spectral weight thus flows from higher to lower energies
so that the difference $\mathrm{Im}(T_{\ge 2}-T_2)$ is expected to become negative at
some energy. We indeed observe this violation of causality in our
calculations but have chosen to accept it,
as the sign change occurs at an energy of several eV, where the lifetime
broadening has already become quite small and where diagrams neglected in the
present work should become more relevant and recover the correct sign. 

\begin{figure}
\includegraphics[width=0.49\columnwidth]{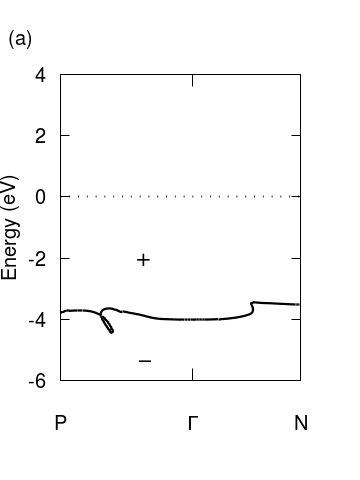}
\includegraphics[width=0.49\columnwidth]{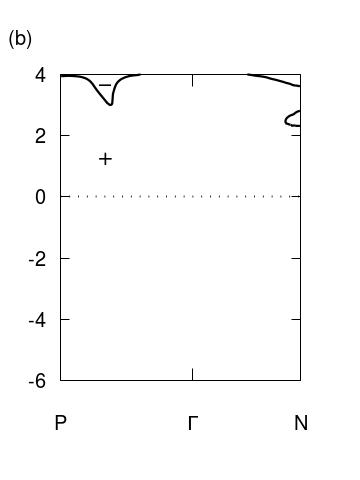}
\caption{\label{fig::causal}
Map of regions where the spectral function $S^\sigma(\mathbf{k},\omega)$
[(a) $\sigma=\uparrow$, (b) $\sigma=\downarrow$] of
iron has positive ($+$) and negative ($-$) sign. The
appearance of the negative sign implies a violation of causality, see text.
The energy zero is set at the Fermi energy.
}
\end{figure}

Figure \ref{fig::causal} shows a map of the sign of the spectral function
$S^\sigma(\mathbf{k},\omega)$ as defined in Eq.~(\ref{Dyson}) for the two
spin channels in the case of iron. The causality condition demands the
spectral function to be non-negative everywhere, but we do see several
regions where the sign is, in fact, negative. 
One notices a spin asymmetry:
the wrong sign appears at negative energies in the spin-up channel, whereas 
it is seen at positive energies in the spin-down channel. As will be explained
in the next section, this spin selectivity is due to the fact that a
coupling to spin waves is possible only for majority holes (occupied states) 
and minority electrons (unoccupied states), which confirms our explanation
that the causality violation originates from a shift of spectral weight from
single-particle Stoner excitations to the low-energy spin waves, leading
to the appearance of the wrong sign in the difference $\mathrm{Im}(T_{\ge 2}-T_2)$. Both
terms individually have the correct sign.
In fact, if we perform the calculation with $T_{\ge 2}$ alone (instead of the
difference), the sign of the spectral function never becomes negative.
However, such a calculation exhibits extreme and unphysical renormalization 
effects.

The lifetime broadenings in the regions with negative sign must be
considered incorrect. These regions are, however, relatively far away from
the Fermi energy, below $-3.5$ eV and above $2.5$ eV for spin up and down, respectively,
and the renormalizations and lifetime broadenings close to the Fermi energy,
where the $GT$ self-energy diagrams dominate, should show the correct behavior.
Clearly, if all infinitely many diagrams were included in the self-energy,
the causality condition should be fulfilled. In this sense, the violation of causality
is an effect of missing self-energy diagrams. For example, for larger
absolute energies it is expected that the imaginary part of the $GW$
diagram, neglected in the present work, becomes more important. In fact,
there is convincing argument that already the $GW$ self-energy term that is second order in $v$
restores the correct sign: This term is formally similar to the self-energy
contribution derived from $T_2$ and $K$ [Eq.~(\ref{TinR})], except that the
multiplication from both sides is with $v$ instead of $W$. Since
$v(\mathbf{r},\mathbf{r}')>W(\mathbf{r},\mathbf{r}';\omega=0)$ for (nearly)
all $\mathbf{r}$ and $\mathbf{r}'$, it seems natural to expect the corresponding
self-energy contribution (also its imaginary part) to be larger than the one
from $T_2$. The combination of
$GT$ and $GW$ is planned as a future study. The question of whether $GW$
fully restores the correct sign must therefore await clarification until then.

}

\section{Calculations} \label{sec::Calculations}


\begin{figure}
\includegraphics[width=1.\columnwidth]{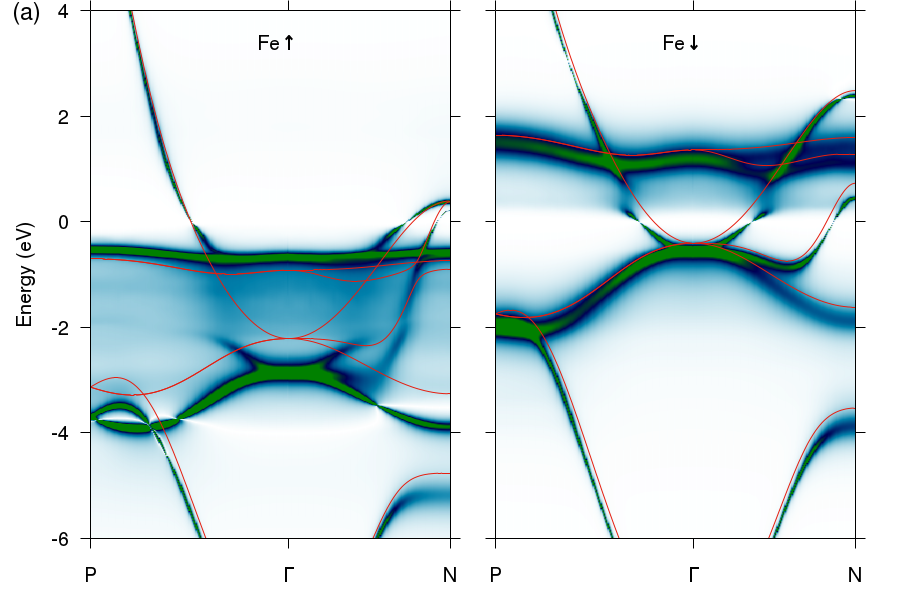}
\includegraphics[width=1.\columnwidth]{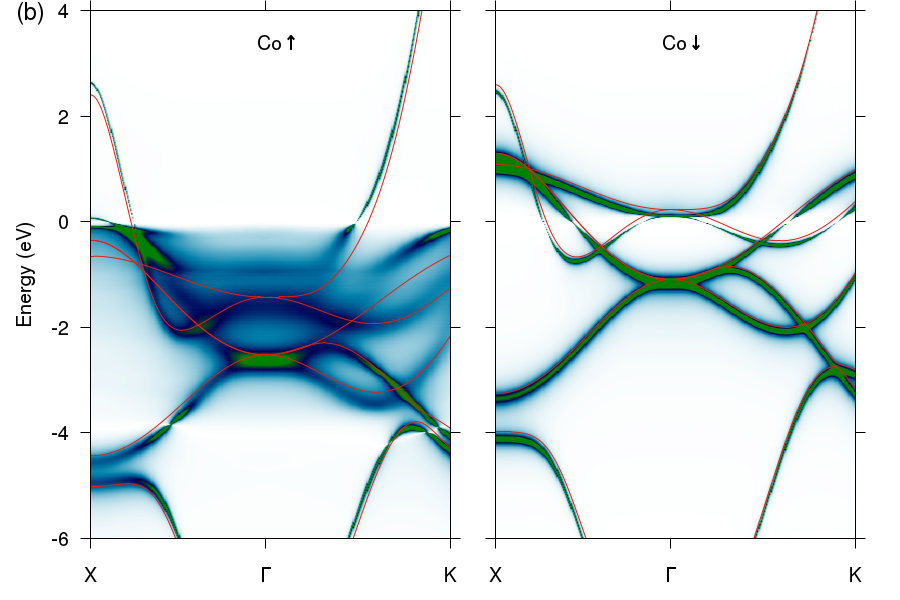}
\includegraphics[width=1.\columnwidth]{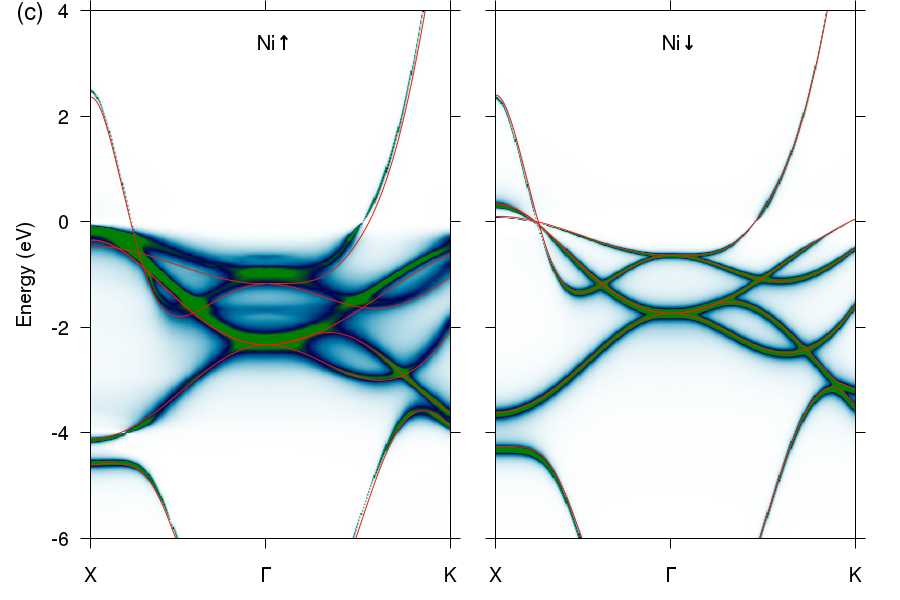}
\caption{\label{fig::bandstr}(Color online) Band structures of (a) Fe, (b) Co,
and (c) Ni. The spectral functions [Eq.~(\ref{Dyson})] are shown in light blue to green. 
The LSDA band structure is shown as red lines for comparison.
The energy zero is set at the Fermi energy. 
The lifetime broadening decreases in the order of increasing
$d$ occupancy (Fe$\rightarrow$Co$\rightarrow$Ni). The spin and particle-hole
asymmetries in the lifetime broadening are clearly seen.
}
\end{figure}

Figure \ref{fig::bandstr} shows the $\mathbf{k}$- and energy-resolved spectral functions
[Eq.~(\ref{Dyson})] in light blue to green together with the LSDA band
structure as red lines for bcc Fe, fcc Co, and fcc Ni. The LSDA band
structure is the uncorrected one (i.e, $\Delta_\mathbf{x}=\Delta_v=0$), while the
non-interacting reference system for the $GT$ renormalization is taken from
an LSDA solution with a parameter $\Delta_\mathrm{x}$ to enforce the
Goldstone condition ($\Delta_\mathrm{x}=0.10$~eV, 0.36~eV, and 0.17~eV for
the three materials, respectively) and a parameter $\Delta_v$ to align the
chemical potentials of the non-interacting and interacting systems 
($\Delta_v=$190~meV, 76~meV, and 25~meV). \cf{The parameters are discussed
in Sec.~\ref{sec::Implementation}.} We employ a Wannier set of $s$, $p$, and
$d$ orbital character constructed from the 18 lowest eigenstates in each
spin channel. The Brillouin zone has been
sampled with a 14$\times$14$\times$14 $\mathbf{k}$-point set.
\cf{The $\mathbf{k}$ integrations of Eq.~(\ref{contdef}) are carried out
using a newly developed tetrahedron method \cite{friedrich19}.

For the first term of Eq.~(\ref{contdef}), we employ a frequency mesh
along the imaginary axis containing 20 points from 0 to $2\mathrm{i}$~htr. A Pad\'{e} approximant,
determined from these 20 points, yields $T_{\mathbf{q}mm'}^{\sigma\sigma'}(\mathbf{k},\mathrm{i}\omega)$ for the   
complete imaginary axis (interpolating between the points and extrapolating
beyond $2\mathrm{i}$~htr) and enables an analytic integration along $\mathrm{i}\omega$. 
For the second term of Eq.~(\ref{contdef}), the matrix
$T_{\mathbf{q}mm'}^{\sigma\sigma'}(\mathbf{k},\omega)$ is calculated for
real frequencies $\omega$ in steps of 20~meV. The self-energy
$\Sigma^\sigma_{\mathbf{q}m}(\omega)$ is evaluated on a mesh between $-6$~eV and
$4$~eV (with respect to the Fermi energy) with a step size of 50~meV. To
achieve very high resolution in energy, necessary to resolve
sharp quasiparticle peaks (as found close to the Fermi energy), 
a spline interpolation is applied to
obtain spectral functions in steps of 2~meV. The calculations
were carried out on the supercomputer JURECA at the J\"ulich
Supercomputing Centre \cite{JSC}.
}

Looking closely at the regions around the Fermi energy,
one realizes that the lifetime broadenings disappear at energy zero and grow rapidly for
lower and higher energies. Beyond a binding energy of around 4~eV
(negative energies in the diagrams) in the spin-up channel, 
they decrease again (even to the extent
that the sign changes to negative, see Sec.~\ref{sec:causality}.)
The reason for this is the restriction of the self-energy to the $GT$
diagram, which is expected to give the low-energy scattering contribution
because of the low energies of the magnon excitations contained in the $T$ matrix.
At higher binding energies, other diagrams, which are neglected in the
present study, should become increasingly important, for example, the $GW$
diagram. 

\begin{figure}
\includegraphics[height=3cm]{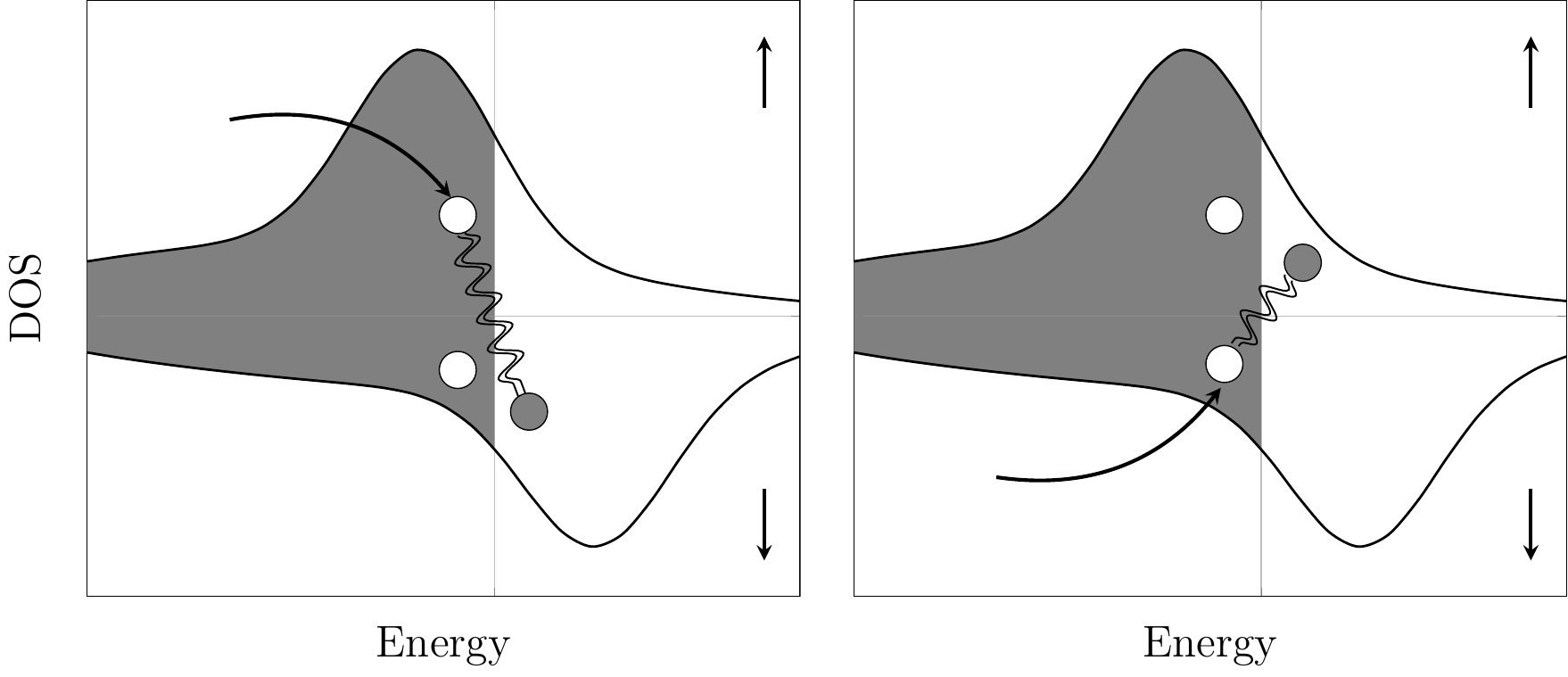}
\caption{\label{schema}Schematic illustration of the scattering of majority
(left) and minority holes (right) in a ferromagnet. The scattering phase space of the former is much
larger than that of the latter.}
\end{figure}

The lifetime broadening is strongest in bcc Fe and weakens in the order of
$d$ occupancy (Fe$\rightarrow$Co$\rightarrow$Ni). This is due to the
decrease of the number of available minority conduction states, into which a majority electron can be
excited, thus reducing the phase space of spin excitations. We also
observe a spin asymmetry of the lifetime effects: the majority valence states are
broadened more strongly than the minority ones. This can be
attributed to the spin asymmetry of the density of states. The scattering diagram
shown in Fig.~\ref{fig::GTSelfEnergy} may be described as an incoming majority hole at 1,
which excites an electron-hole pair at 2 in the minority channel. The
majority hole and minority electron then propagate through the system,
successively scattering with each other until the minority electron-hole
pair recombines at 4. The correlated propagation of the electron and the
hole (having opposite spins) 
embodies the spin-flip excitations in the form of Stoner excitations and,
due to their correlated motion, also of the collective spin-wave excitations. 
The probability for the pair formation and the pair propagation process depends on
the number of available majority valence and minority conduction states,
which is large in this case, see Fig.~\ref{schema}, left. 
(We note that the states also have similar orbital character.)
On the other hand, an incoming minority hole would have to combine with
majority conduction states. Naturally, the available phase space is much
smaller in this case (Fig.~\ref{schema}, right), which explains the spin asymmetry in the lifetime
broadening.
The fact that minority bands are visible down to much larger binding
energies than majority bands in ARPES is thus revealed to be a
lifetime effect caused by electron-electron scattering.
It should be noted that the qualitative behavior has already
been observed in previous studies based on DMFT
\cite{katsnelson99,grechnev07,sanchez-barriga09,sanchez-barriga12} and model calculations
\cite{mills99}.
Our calculated spectral functions for iron are remarkably
similar to the ones published in Ref.~\onlinecite{sanchez-barriga12}. However,
the fact that both the magnon excitation spectrum and the $GT$ self-energy
are fully $\mathbf{k}$ dependent
in our method gives rise to
additional fine structure in the spectral functions, e.g., band anomalies,
which have not been observed in theoretical studies so far.

Apart from the spin asymmetry, there is evidence for a particle-hole asymmetry as well, i.e., 
a markedly different lifetime broadening above and below
the Fermi energy; namely, in the majority channels of all three materials and also in the
minority channel of Fe. Here, an explanation very similar to the
one just given would predict the lifetime broadening of majority hole
(minority electron) bands to be larger than the broadening of majority
electron (minority hole) bands, in agreement with the spectra.

In iron, and to a lesser degree in Co,
the many-body renormalization is so strong that 
the quasiparticle character is virtually lost between
$-2.5$ and $-0.5$~eV in the majority channel. A similar but less pronounced
renormalization effect is also seen in the minority channel of iron at around 0.5~eV
above the Fermi energy, see below.
The $GT$ self-energy thus leads to 
a substantially more complex spectral function than the quasiparticle band structure of a
$GW$ calculation, let alone the mean-field band structure of Kohn-Sham DFT.
In the case of
the former, one often only plots the energy-renormalized quasiparticle bands
and drops the lifetime broadening altogether, which is legitimate because the
main self-energy poles of $GW$ are typically energetically far away from the 
valence and low-lying conduction bands. The quasiparticle approximation is
therefore a good one. To be more precise, the main $GW$ self-energy
poles describe the coupling of the electrons and holes to plasmons (emission
and absorption of plasmons), whose energies $\approx \epsilon_\mathrm{P}$
are typically in the order of tens of eV. The poles, thus, appear well
separated from the valence (conduction) band energies
$\epsilon^\sigma_{\mathbf{k}m}$ at around
$\epsilon^\sigma_{\mathbf{k}m}\mp\epsilon_\mathrm{P}$ for valence and
conduction states, respectively. As a result, one obtains
a well-defined renormalized quasiparticle band structure and, at a much larger energy,
the corresponding plasmon satellites. In the $GT$ self-energy, on the other
hand, the plasmon energies $\epsilon_\mathrm{P}$ would have to be replaced by 
the much smaller magnon energies $\approx \epsilon_\mathrm{M}$. Furthermore,
the spin of the Green function in the $GT$ self-energy is opposite to that of the self-energy itself [see
Eq.~(\ref{eq::ElMagSelfEnergy})]. So, the $GT$ self-energy poles are to be
expected at energies shifted from the \emph{opposite-spin} band energies,
i.e., at around $\epsilon^{-\sigma}_{\mathbf{k}m}\mp\epsilon_\mathrm{M}$. 
In this sense, the magnon energies additionally compete with the exchange splitting.
To sum up, the $GT$ self-energy poles will fall straight into the valence 
(and low-lying conduction) region, 
giving rise to a much richer photoelectron spectrum than, for example, in $GW$ calculations.

The plot of the $\mathbf{k}$- and energy-resolved spectral function allows fine details
of the quasiparticle band structure to be investigated. The resonant
interaction with many-body states in the valence band region (these form the
pole structure of the self-energy)
can give rise to anomalies in the band dispersion and satellite bands, 
as we will discuss exemplarily for a few cases in the following.

\cf{

\begin{figure}
\includegraphics[width=1.\columnwidth]{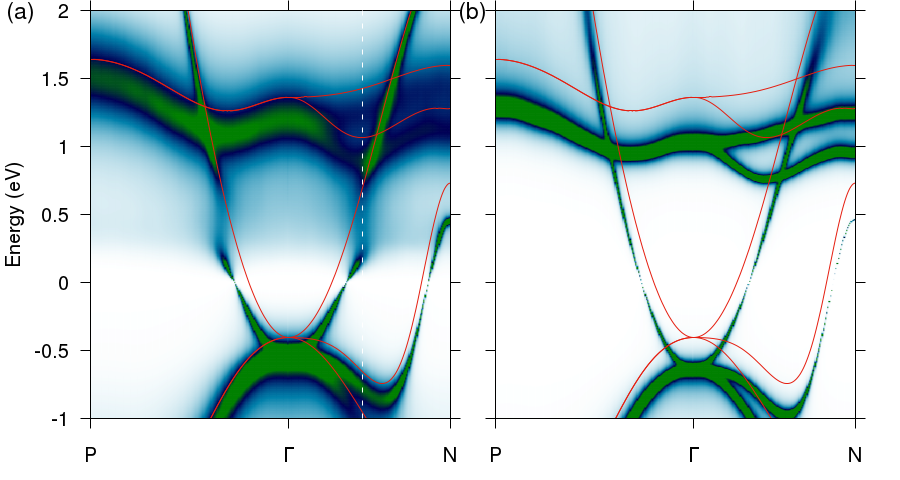}
\includegraphics[width=0.5\columnwidth,angle=-90]{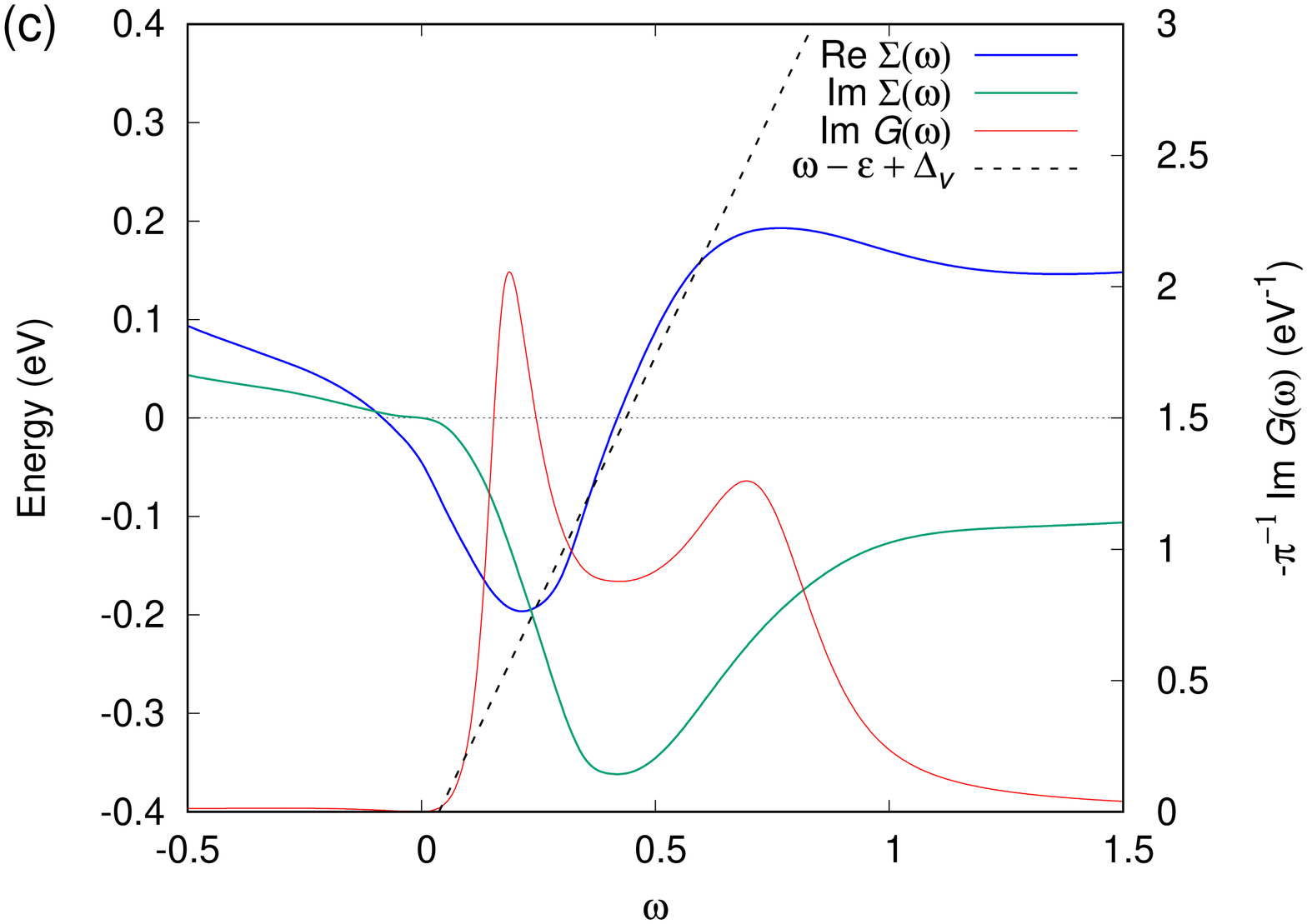}
\caption{\label{fig::anomaly}
(Color online)
(a) Closeup of dispersion anomalies in a spin-down conduction band of iron.
The LSDA bands are shown in red for comparison.
(b) Same band renormalized only with the lowest-order diagram, which is of 
third order in $W$, see Fig.~\ref{fig::SelfEnergyExpansion}, and which cannot
describe the scattering of electrons and spin waves. The
anomaly disappears. The anomaly in (a) can therefore be attributed to
electron--spin-wave scattering. (c) Spectral function (red line) along the white
dashed line marked in (a). Real and imaginary parts of the self-energy are
shown as blue and green lines, respectively. The spectral-function peaks are
roughly at the intersections of the blue curve with a linear function, see
text.
}
\end{figure}

We first analyze a band anomaly in a spin-down conduction band of iron
about 500~meV above the Fermi energy, see Fig.~\ref{fig::bandstr}(a), right
panel. A closeup is shown in Fig.~\ref{fig::anomaly}(a). Whereas the
respective Kohn-Sham band (red lines) is continuous and shows an
approximately parabolic dispersion, the renormalized quasiparticle band
seems to be cut in two: the band breaks off at around 250~meV and reappears at
700~meV with an incoherent continuum in-between. The energetic position of
the anomaly, immediately above the Fermi energy, indicates that electronic 
scattering with low-energy spin waves is responsible for its formation. 
This is confirmed by Fig.~\ref{fig::anomaly}(b), which shows the band 
structure renormalized with only the first term of Eq.~(\ref{eq::TmatrixBSE}).
This term, the third-order diagram of Fig.~\ref{fig::GTSelfEnergy}, can be
regarded as not containing scattering contributions with collective spin-wave
excitations, which, in turn, would require an infinite (or at least large) number of ladder 
diagrams to be summed to become a well defined many-body excitation. 
Fig.~\ref{fig::anomaly}(b) does not exhibit any band anomaly in the
respective bands \footnote{The relatively strong energetic shift results from employing the
same parameters $\Delta_\mathrm{x}$ and $\Delta_v$ as for the full $GT$
calculations, without readjusting them to the self-energy presently
restricted to the third-order diagram.}.
This proves that the scattering with spin waves is responsible for the
appearance of the band anomaly. It also explains why DMFT cannot reproduce
the anomaly because DMFT spatially confines the bosonic excitations in the
impurity, which prevents the spin waves to become well-defined extended many-body 
excitations, despite the fact that infinitely many ladder diagrams are
summed in principle. The spatial confinement in DMFT is equivalent to the
missing $\mathbf{k}$ dependence of the bosonic excitation spectrum. 
In contrast, this $\mathbf{k}$ dependence (and also the one for the
self-energy) is fully preserved in our theoretical approach.

The electron-magnon interaction gives rise to a double-peak structure in the
spectral function at the Bloch momentum of the anomaly [white dashed line in
Fig.~\ref{fig::anomaly}(a)]. Figure \ref{fig::anomaly}(c) shows this
spectral function as the red curve. The blue and green lines show the real
and imaginary components of the self-energy matrix element
$\Sigma^\sigma_{\mathbf{k}m}(\omega)=\langle\varphi^\sigma_{\mathbf{k}m}|\Sigma^\sigma(\omega)|\varphi^\sigma_{\mathbf{k}m}\rangle$,
respectively, where $\varphi^\sigma_{\mathbf{k}m}$ is the respective
Kohn-Sham state. The spectral function clearly exhibits two peaks. In
general, peaks in the spectral function are expected where the denominator
of Eq.~(\ref{Dyson}) becomes small, i.e., where the linear function
$\omega-\epsilon^\sigma_{\mathbf{k}m}+\Delta_v$ becomes identical or, at
least, close to the self-energy $\Sigma^\sigma_{\mathbf{k}m}(\omega)$. In
fact, the two peaks are seen close to the intersections of the linear
function (black dashed line) with the real part of the self-energy (blue line), 
whereas its imaginary part slightly shifts the peak positions and also
suppresses a possible third peak
related to the intersection in the middle. [If the self-energy is assumed to
be given by a single pole $(\omega-\omega_0)^{-1}$ with a complex
self-energy resonance at $\omega_0$, then equating with the linear function would
yield a quadratic equation with only two possible solutions.]

One might be tempted to identify one of the two peaks as the quasiparticle
peak and the other as the magnon satellite, roughly corresponding to the two
energies $\epsilon^\sigma_{\mathbf{k}m}$ and
$\epsilon^{-\sigma}_{\mathbf{k}m}+\epsilon_{\mathrm{M}}$
discussed above. However, this is not quite correct because it would
disregard the coupling between these two fundamental many-body states. Let us
consider a simple $2\times 2$ Hamiltonian with the above two energies on the
diagonal. The offdiagonal (coupling) matrix elements must be nonzero,
otherwise there would not be any renormalization. As the quasiparticle band
$\epsilon^\sigma_{\mathbf{k}m}$ disperses upwards in energy, it approaches
and transverses the energy
$\epsilon^{-\sigma}_{\mathbf{k}m}+\epsilon_{\mathrm{M}}$. The coupling and
the strong lifetime effects make the band disappear. When the
distance between the two energies has again grown large enough, the quasiparticle band reappears. The
two peaks show up when the diagonal elements are roughly identical. The peak
separation can then be identified as a measure for the electron-magnon
coupling strength (offdiagonal elements).

\begin{figure}
(a)\includegraphics[width=0.5\columnwidth,angle=-90]{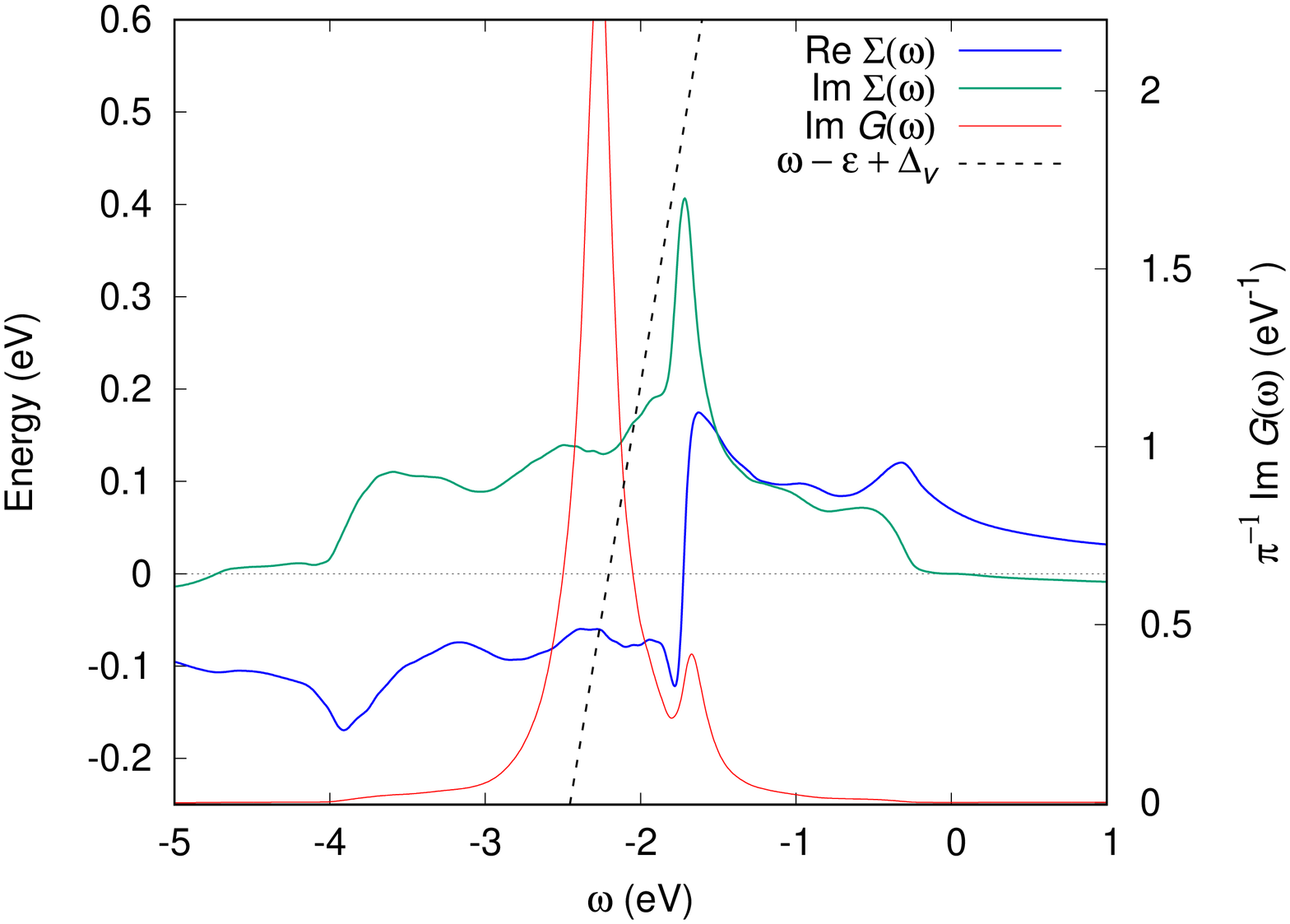}
(b)\includegraphics[width=0.5\columnwidth,angle=-90]{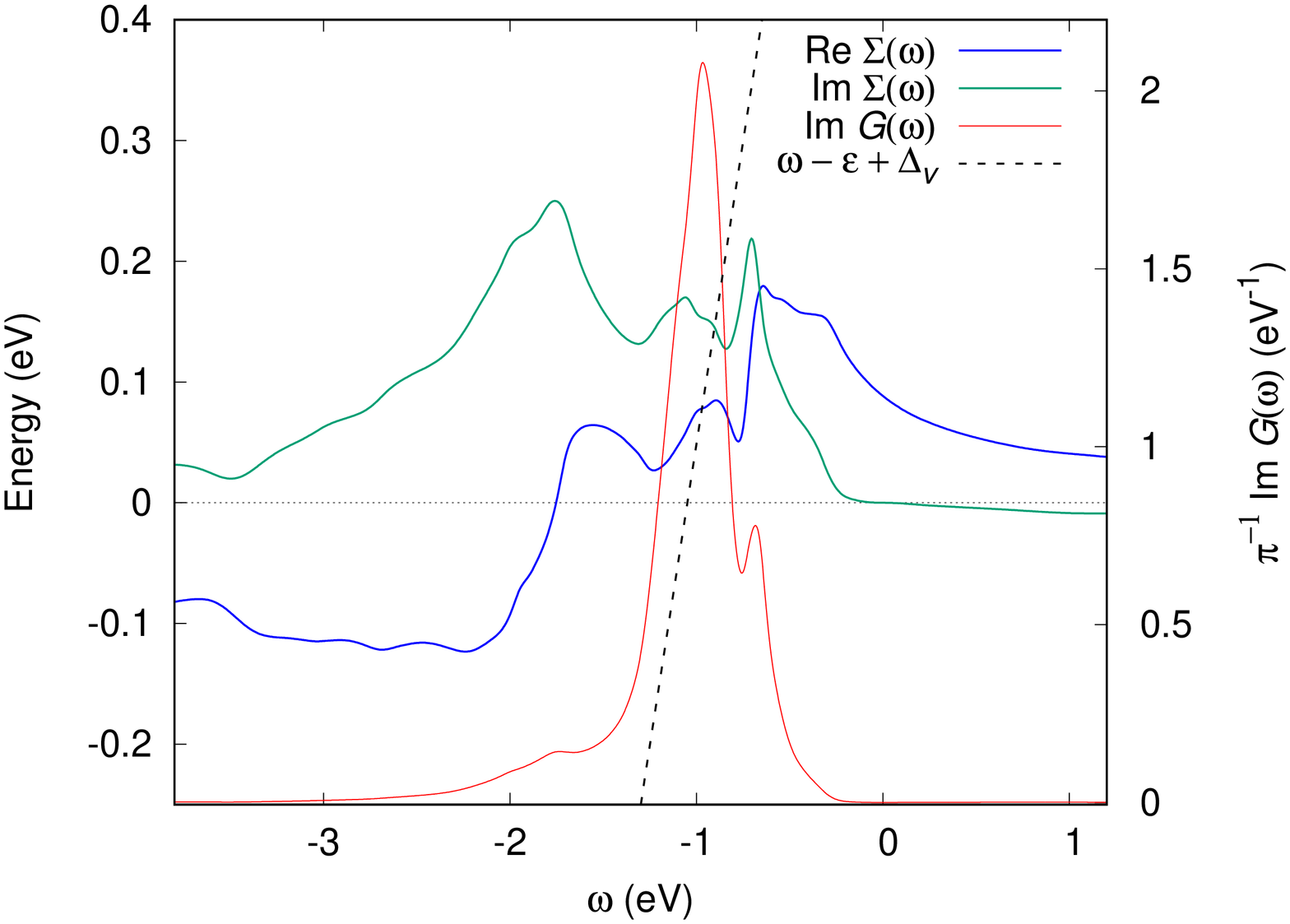}
\caption{\label{fig::magnaron}
(Color online) Same as Fig.~\ref{fig::anomaly}(c) for the two secondary bands [(a) upper
one, (b) lower one] seen in the
spin-up channel of nickel of Fig.~\ref{fig::bandstr}(c). The spectral
function and self-energy are plotted for the $\Gamma$ point.
}
\end{figure}

In the renormalized majority band structure of nickel,
Fig.~\ref{fig::bandstr}(c), one observes two curious \emph{secondary} bands
close to the center of the Brillouin zone, a weak one above the highest
valence band and a somewhat stronger one between the highest and the second
highest band. The latter one even seems to trace the dispersion of the
quasiparticle bands below. Although the features look
very different to the band anomaly discussed before, they have a similar
origin. One might imagine the two energies $\epsilon^\sigma_{\mathbf{k}m}$ and
$\epsilon^{-\sigma}_{\mathbf{k}m}-\epsilon_{\mathrm{M}}$ to be nearly
constant with respect to the Bloch vector $\mathbf{k}$. The two peaks,
the strong quasiparticle and the weak secondary one, then form two nearly
parallel bands with little variation in strength. Figs.~\ref{fig::magnaron}(a) and (b)
show the spectral function, together with the self-energy and the linear
function $\omega-\epsilon^\sigma_{\mathbf{k}m}+\Delta_v$, at the $\Gamma$
point ($k=0$). In contrast to Fig.~\ref{fig::anomaly}(c), the two spectral-function
peaks are of very different strength. The small secondary peak cannot be
related to an intersection of the black dashed line with the blue solid one.
It rather appears where the two functions have the closest approach.
Due to the very different peak strengths, one can interpret the strong peak
as the quasiparticle peak and the secondary one as a magnon satellite peak
(the corresponding many-body states having small admixtures of each other).
It might seem a curiosity that the satellite peak appears \emph{above}
the quasiparticle peak instead below, as one is used to from $GW$
calculations. This can be attributed to the exchange splitting, which, as
discussed above, competes with the magnon energy and can cause the satellite
peak to emerge \emph{on the wrong side}, according to 
$\epsilon^{-\sigma}_{\mathbf{k}m}+\epsilon_{\mathrm{M}}
=\epsilon^{\sigma}_{\mathbf{k}m}-\epsilon_{\mathrm{M}}+\Delta_{\mathbf{k}m}$
with the exchange splitting $\Delta_{\mathbf{k}m}$.

}

In iron and cobalt, we clearly see the onset of strong lifetime
broadening when the quasiparticle bands suddenly lose intensity, and a
``waterfall'' structure forms. 
The ``waterfall'' structures seen here are very similar to
a feature often observed in ARPES spectroscopy. This feature is
interpreted as a footprint of a many-body resonance, at whose energy the
spectral function is expected to show non-Fermi liquid behavior. The
interpretation is confirmed by the present calculations.

\begin{figure}
\includegraphics[width=1.\columnwidth]{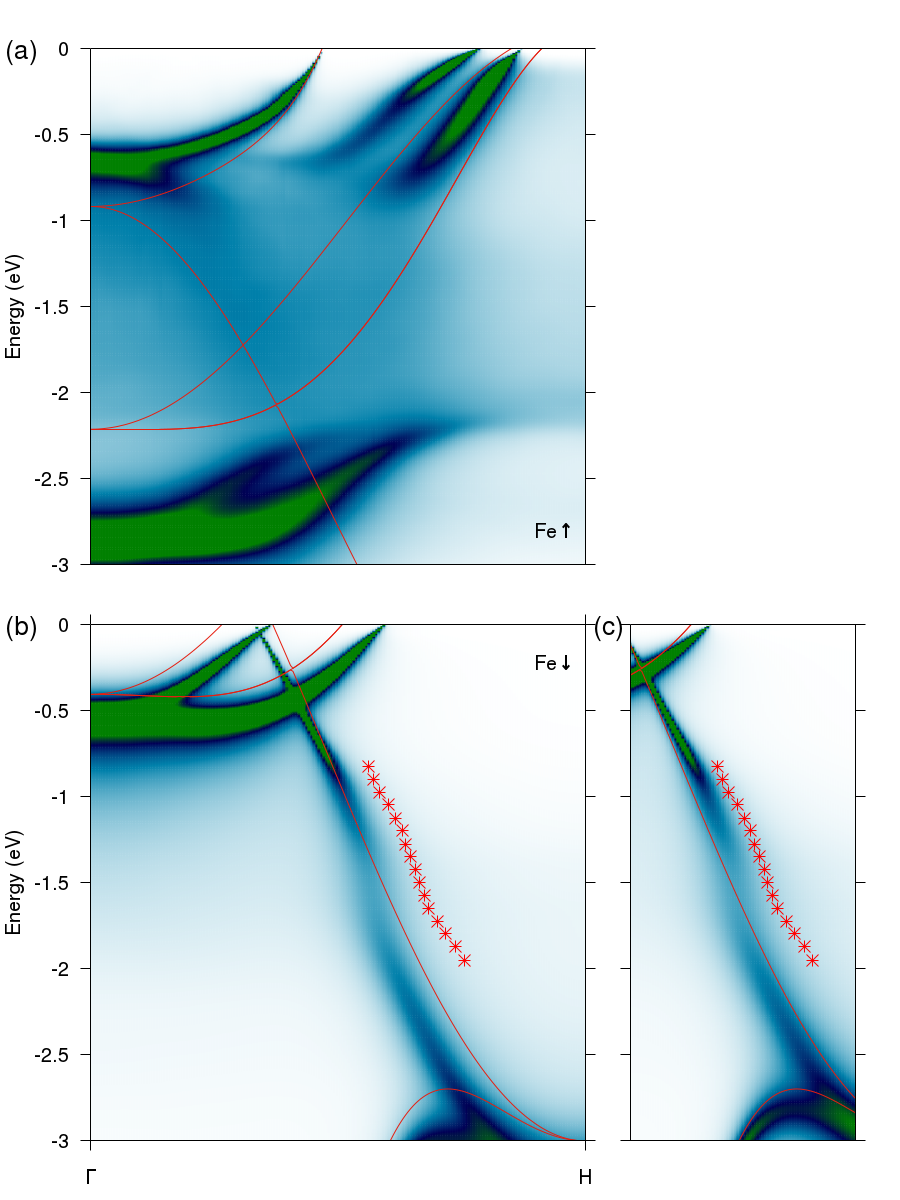}
\caption{\label{fig::kinks2}(Color online) Renormalized band structure of iron in the (a)
majority and (b) minority channel. High-energy anomaly in a minority band of iron.
(a) and (b) $GT$-renormalized spectral function with $\Delta_v=190$~meV and (c)
$\Delta_v=0$. The LSDA bands are shown as red lines. The red crosses are fitted
peak positions from experiment (data reproduced from
Ref.~\onlinecite{Mlynczak19}).
}
\end{figure}

The ``waterfall'' structures are observed roughly at 250-600~meV in Fe [Fig.~\ref{fig::kinks2}(a)] and at 200-500~meV
in Co [Fig.~\ref{fig::bandstr}(b)]. 
Nickel, on the other hand, exhibits the weakest lifetime broadening of all. While it is still strong, it does not lead to the
disappearance of the quasiparticle character as in Fe and Co. The
quasiparticle bands, although substantially broadened, remain well-defined over the whole energy range. 
Nevertheless, one can still make out a slight waterfall shape in the
strongly dispersing band between $\Gamma$ and K.
\cf{In one of the bands of Fe, one can discern a slight kink in the band
dispersion at about 250~meV, just where the lifetime broadening sets in. 
Cui \textit{et al.}~\cite{cui07} published ARPES spectra
of iron and claimed an anomaly in a majority band at around 270~meV to originate from
electron-magnon interactions. However, in Ref.~\onlinecite{cui10} they
distanced themselves from this claim and attributed the anomaly to an
inaccuracy of the line-shape analysis in case of strong lifetime broadening. 
Indeed, our theoretical spectra do not exhibit a dispersion anomaly in the
band they have measured.}

In the following, we investigate another band
anomaly in a minority band of iron at the relatively large binding energy of
1.5~eV. Interestingly, very recently, an anomaly was found in ARPES
experiments in the same minority band and roughly at the same energy and momentum as
predicted by our theoretical calculation. We report about this comparison elsewhere (Ref.
\onlinecite{Mlynczak19}) in detail. 
Figure~\ref{fig::kinks2}
(b) and (c) present the theoretical (minority) spectral functions together with a
comparison to the peak fits (red crosses) of the experimental momentum distribution curves. The
two calculations (b) and (c) differ in the parameter $\Delta_v=190$~meV and
$\Delta_v=0$. 
The form of the two plots is very similar,
implying that the spectral function is largely unaffected by the parameter $\Delta_v$
at higher binding energies.
In Ref.~\onlinecite{Mlynczak19},
we only show the spectrum for $\Delta_v=0$, which is why we have included this
result here for comparison. 

After what we have discussed before, it might seem contradictory that a
band anomaly should appear in a minority valence band. In fact, the
scattering phase space is smaller for the minority valence states, but it is
not zero. It does contain spin-flip Stoner excitations to which a
propagating hole can couple. And, in the present case, these Stoner
excitations form a particularly strong intensity in the spin excitation
spectrum at around 0.7~eV \cite{Mlynczak19}. The many-body resonance is then formed by the Stoner
pairs (spin $+\hbar$) together with majority hole
$d$ states (spin $-\hbar/2$) that form a pronounced peak in the
density of states at around 0.8~eV. This resonance is strong enough to give rise to the
observed anomaly in the minority hole band (spin $+\hbar/2$) at a binding
energy of 1.5~eV. We note at this
point that the shape of the band anomaly can again be described as a
``waterfall'' structure.


\begin{figure}
\includegraphics[width=\columnwidth]{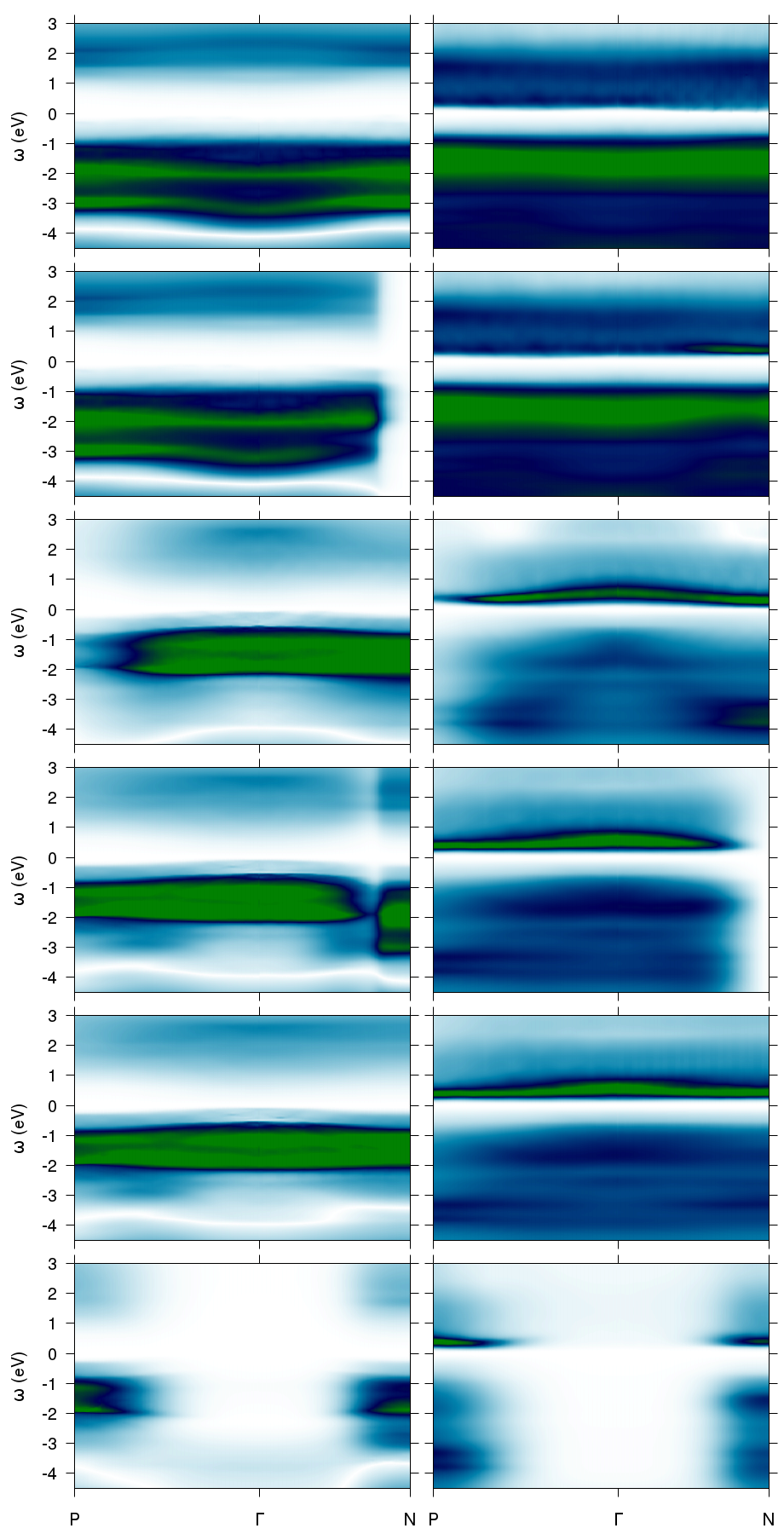}
\caption{\label{fig::senergy}(Color online) Momentum and frequency dependence of the 
self-energy for iron. The imaginary component $|\mathrm{Im}\,\Sigma^\sigma_{\mathbf{q}m}(\omega)|$ 
is shown for the bands marked in color in Fig.~\ref{fig::bandk},
from bottom to top: green, blue, red, magenta, orange, black; left column
spin up, right column spin down.
}
\end{figure}

\begin{figure}
\includegraphics[width=\columnwidth]{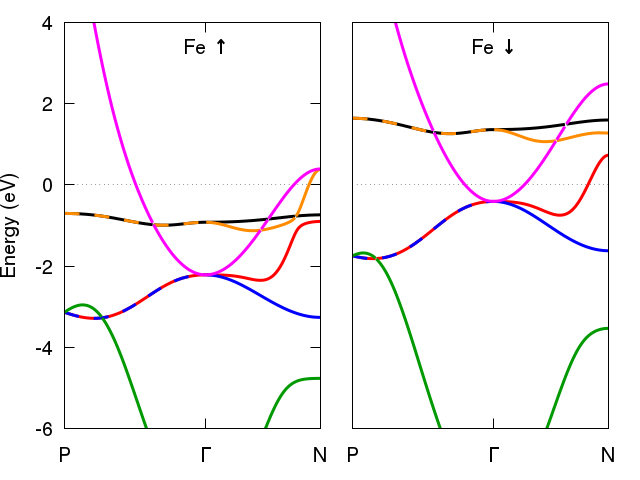}
\caption{\label{fig::bandk}(Color online) Kohn-Sham LSDA band structure of iron. The
colors mark the different bands for which the self-energy is plotted in
Fig.~\ref{fig::senergy}.}
\end{figure}

\cf{

The present method fully retains the $\mathbf{k}$ dependence of both the bosonic
(magnonic) excitation spectrum and the self-energy. This is in contrast to the DMFT
approach, in which the spatial confinement to the impurity effectively
obliterates the $\mathbf{k}$ dependence of the quantities. This leads to the
fact that band anomalies that are caused by a coupling to extended spin
waves are properly described in the present method but might be missed in
DMFT, see, e.g., Fig.~\ref{fig::anomaly}(a). The $\mathbf{k}$ dependence of
the spin excitation spectrum has been described already in
Refs.~\cite{sasioglu10,friedrich14,friedrich18,mueller16} among other works. However, we
also have the possibility to directly investigate the $\mathbf{k}$
dependence of the self-energy. To this end, we have plotted the self-energy
matrix elements $\Sigma^\sigma_{\mathbf{k}m}(\omega)$ in
Fig.~\ref{fig::senergy}. The respective bands are marked in color in
Fig.~\ref{fig::bandk}. Overall, we observe a relatively weak dependence of the
self-energy on the Bloch momentum, in particular for the localized $d$
bands. However, some bands do show a strong $\mathbf{k}$ dependence, especially
the more strongly dispersing bands, such as the green band
(Fig.~\ref{fig::bandk}).
We also see the effect of hybridization: For example, the spin-up red and orange
curves strongly hybridize close to the N point and exchange orbital
character. This is reflected in the self-energies, which, towards the right,
seem to exchange roles. Again, we would like to emphasize that a weak
$\mathbf{k}$ dependence of the self-energy does not necessarily speak in favor 
of a purely local treatment of correlation effects. Equally important, or arguably
more important, is the $\mathbf{k}$ dependence of the bosonic excitation
spectrum. A purely local treatment would turn the quadratic spin-wave branch
into an incoherent broad peak, which might wash out resonances of the
electron-magnon scattering.

}

\section{Conclusions} \label{sec::Conclusions}

In summary, we have described a Green-function technique to calculate the
electron-magnon scattering from first principles. The $GT$ self-energy is
constructed from the product of the single-particle Green function $G$ and the
multiple scattering $T$ matrix, which can be viewed as an effective
interaction that acts through the exchange of magnons. The $T$ matrix, thus,
contains the full spin excitation spectrum, comprising collective spin-wave
and single-particle Stoner excitations, and, diagrammatically, it
describes the correlated motion of an electron-hole pair of opposite spins.
When expanded in powers of the screened Coulomb interaction $W$, the $T$
matrix can be written as a summation over ladder diagrams of ever increasing
order. We have shown that a formulation consistent with the Hedin equations
requires the $GT$ self-energy to include the third-order ladder diagram as
the lowest order. In this way, the self-energy approximation is free of
double counting with the Hartree or exchange ($GW$) diagrams. 

\cf{
We showed that the so-constructed $GT$ self-energy can lead to a violation of 
causality, manifesting itself in an incorrect sign change of the spectral
function. We elucidated the mathematical reason for this violation of
causality: The particular subset of $GT$ diagrams can be written as the
difference of two terms, respectively containing the renormalized and the bare
electron-hole propagator. The renormalization redistributes the spectral
weights in such a way that the difference can become negative in a certain
energy region. However, since the sign change occurs far away from the Fermi
energy, we accept this violation of causality and show that the resulting
renormalized band structures yield a correct description of the
electron-magnon scattering around the Fermi energy. It is expected that
other diagrams neglected in the present work, in particular, the $GW$ self-energy, 
will act to restore the correct sign of the spectral function. However, this
goes beyond the scope of the present study.
}

The $GT$ self-energy gives rise to a very strong lifetime broadening of the
quasiparticle bands, to the extent that in a certain energy region the
quasiparticle character is virtually lost in the majority valence bands of
bcc Fe and fcc Co. The minority bands, however, are much less affected by
the lifetime broadening, which we have attributed to the spin asymmetry of the
density of states. For a similar reason, one also finds a particle-hole
asymmetry. The spin asymmetry in the lifetime broadening explains the
experimental fact that minority bands are seen to relatively large binding
energies, while the majority bands disappear very soon below the Fermi energy.
The strong renormalization effects also give rise to ``waterfall''
structures in the $\mathbf{k}$- and energy-resolved spectral functions, in
particular for bcc Fe and fcc Co.

The $T$ matrix contains the acoustic spin-wave excitations for all
wave vectors, including the long-wavelength limit
$\mathbf{k}\rightarrow 0$, where the excitation energy vanishes and the $T$
matrix becomes singular with a mathematical delta peak in its imaginary part.
Since such a singularity is
likely to give rise to numerical difficulties, we have investigated its
contribution to the $GT$ self-energy. As a result, the contribution of the
long-wavelength spin-wave excitation should be very small in the problematic limit 
$\mathbf{k}\rightarrow 0$. 

The contribution of the spin-wave excitations do not vanish for finite $k$,
though. Our numerical calculations have shown that it is this contribution
\cf{that can lead to the formation of dispersion anomalies in majority valence and
minority conduction bands. There, the spectral function can exhibit double-peak 
structures, which may be interpreted as a quasiparticle and a magnon satellite
peak. However, when the peaks are of equal strength, they are general mixtures of
the two and the peak separation can be regarded as a measure of the electron-magnon coupling
strength. }


Our method can describe a band-dispersion anomaly in the minority
channel of bcc Fe at a large binding energy of about 1.5~eV, which has very
recently been observed in ARPES measurements \cite{Mlynczak19}. Whereas such a
large energy is commonly not associated with a magnon excitation, we could
elucidate \cite{Mlynczak19} that this band anomaly is caused by a many-body resonance formed
by a superposition of Stoner excitations (an increased Stoner intensity is
seen in the spin excitation spectrum at around $0.7$~eV) and majority hole 
states that form a peak in the density of states (at around $0.8$~eV).

\cf{

While the overall shape of the renormalized band structures is similar to 
corresponding DMFT calculations, the latter have not been able to describe
the band anomalies discussed in the present paper. We attribute this to the
main approximation of the DMFT approach, the spatial confinement of the
electronic correlation to a single impurity site. This neglects the $\mathbf{k}$
dependence of the self-energy, but, more importantly, it also does not allow for a
$\mathbf{k}$ dependent bosonic (magnonic) excitation spectrum, effectively
replacing the well-defined quadratic branch of extended spin waves by a
broad feature of local excitations that cannot couple strongly to electrons.

}

In summary, we believe that the present work is an important step in the
development of ab initio many-body methods for the treatment of magnetic
materials, in which the magnon energies (collective and Stoner
type), the exchange splitting, the band energies, and the electron-magnon coupling strength
set different energy scales that compete with each other. Further studies in
this direction might also elucidate the role of magnons in the potential formation of
Cooper pairs in high-temperature superconductors.

\begin{acknowledgments}
The authors acknowledge valuable discussions with
Ewa M{\l}y\'{n}czak, Lukas Plucinski, Ersoy \c{S}a\c{s}{\i}o\u{g}lu, Markus Betzinger, Manuel dos Santos Dias, Ferdi Aryasetiawan, and Hans Lustfeld.
We thank Irene Aguilera for a critical reading of the manuscript.
We gratefully acknowledge the computing time granted through JARA-HPC on the
supercomputer JURECA at Forschungszentrum J\"ulich.
\end{acknowledgments}


\newpage
\begin{thebibliography}{61}
\expandafter\ifx\csname natexlab\endcsname\relax\def\natexlab#1{#1}\fi
\expandafter\ifx\csname bibnamefont\endcsname\relax
  \def\bibnamefont#1{#1}\fi
\expandafter\ifx\csname bibfnamefont\endcsname\relax
  \def\bibfnamefont#1{#1}\fi
\expandafter\ifx\csname citenamefont\endcsname\relax
  \def\citenamefont#1{#1}\fi
\expandafter\ifx\csname url\endcsname\relax
  \def\url#1{\texttt{#1}}\fi
\expandafter\ifx\csname urlprefix\endcsname\relax\def\urlprefix{URL }\fi
\providecommand{\bibinfo}[2]{#2}
\providecommand{\eprint}[2][]{\url{#2}}

\bibitem[{\citenamefont{\ifmmode \check{Z}\else
  \v{Z}\fi{}uti\ifmmode~\acute{c}\else \'{c}\fi{}
  et~al.}(2004)\citenamefont{\ifmmode \check{Z}\else
  \v{Z}\fi{}uti\ifmmode~\acute{c}\else \'{c}\fi{}, Fabian, and
  Das~Sarma}}]{zutic04}
\bibinfo{author}{\bibfnamefont{I.}~\bibnamefont{\ifmmode \check{Z}\else
  \v{Z}\fi{}uti\ifmmode~\acute{c}\else \'{c}\fi{}}},
  \bibinfo{author}{\bibfnamefont{J.}~\bibnamefont{Fabian}}, \bibnamefont{and}
  \bibinfo{author}{\bibfnamefont{S.}~\bibnamefont{Das~Sarma}},
  \bibinfo{journal}{Rev. Mod. Phys.} \textbf{\bibinfo{volume}{76}},
  \bibinfo{pages}{323} (\bibinfo{year}{2004}).

\bibitem[{\citenamefont{MacDonald et~al.}(1998)\citenamefont{MacDonald,
  Jungwirth, and Kasner}}]{macdonald98}
\bibinfo{author}{\bibfnamefont{A.~H.} \bibnamefont{MacDonald}},
  \bibinfo{author}{\bibfnamefont{T.}~\bibnamefont{Jungwirth}},
  \bibnamefont{and} \bibinfo{author}{\bibfnamefont{M.}~\bibnamefont{Kasner}},
  \bibinfo{journal}{Phys. Rev. Lett.} \textbf{\bibinfo{volume}{81}},
  \bibinfo{pages}{705} (\bibinfo{year}{1998}).

\bibitem[{\citenamefont{Balashov et~al.}(2006)\citenamefont{Balashov, Tak\'acs,
  Wulfhekel, and Kirschner}}]{balashov06}
\bibinfo{author}{\bibfnamefont{T.}~\bibnamefont{Balashov}},
  \bibinfo{author}{\bibfnamefont{A.~F.} \bibnamefont{Tak\'acs}},
  \bibinfo{author}{\bibfnamefont{W.}~\bibnamefont{Wulfhekel}},
  \bibnamefont{and}
  \bibinfo{author}{\bibfnamefont{J.}~\bibnamefont{Kirschner}},
  \bibinfo{journal}{Phys. Rev. Lett.} \textbf{\bibinfo{volume}{97}},
  \bibinfo{pages}{187201} (\bibinfo{year}{2006}).

\bibitem[{\citenamefont{Balashov et~al.}(2008)\citenamefont{Balashov, Tak\'acs,
  D\"ane, Ernst, Bruno, and Wulfhekel}}]{balashov08}
\bibinfo{author}{\bibfnamefont{T.}~\bibnamefont{Balashov}},
  \bibinfo{author}{\bibfnamefont{A.~F.} \bibnamefont{Tak\'acs}},
  \bibinfo{author}{\bibfnamefont{M.}~\bibnamefont{D\"ane}},
  \bibinfo{author}{\bibfnamefont{A.}~\bibnamefont{Ernst}},
  \bibinfo{author}{\bibfnamefont{P.}~\bibnamefont{Bruno}}, \bibnamefont{and}
  \bibinfo{author}{\bibfnamefont{W.}~\bibnamefont{Wulfhekel}},
  \bibinfo{journal}{Phys. Rev. B} \textbf{\bibinfo{volume}{78}},
  \bibinfo{pages}{174404} (\bibinfo{year}{2008}).

\bibitem[{\citenamefont{Schweflinghaus
  et~al.}(2014)\citenamefont{Schweflinghaus, dos Santos~Dias, Costa, and
  Lounis}}]{schweflinghaus14}
\bibinfo{author}{\bibfnamefont{B.}~\bibnamefont{Schweflinghaus}},
  \bibinfo{author}{\bibfnamefont{M.}~\bibnamefont{dos Santos~Dias}},
  \bibinfo{author}{\bibfnamefont{A.~T.} \bibnamefont{Costa}}, \bibnamefont{and}
  \bibinfo{author}{\bibfnamefont{S.}~\bibnamefont{Lounis}},
  \bibinfo{journal}{Phys. Rev. B} \textbf{\bibinfo{volume}{89}},
  \bibinfo{pages}{235439} (\bibinfo{year}{2014}).

\bibitem[{\citenamefont{Schweflinghaus}(2016)}]{schweflinghaus16}
\bibinfo{author}{\bibfnamefont{B.}~\bibnamefont{Schweflinghaus}}, Ph.D. thesis,
  \bibinfo{school}{RWTH Aachen} (\bibinfo{year}{2016}).

\bibitem[{\citenamefont{Dagotto}(1994)}]{dagotto94}
\bibinfo{author}{\bibfnamefont{E.}~\bibnamefont{Dagotto}},
  \bibinfo{journal}{Rev. Mod. Phys.} \textbf{\bibinfo{volume}{66}},
  \bibinfo{pages}{763} (\bibinfo{year}{1994}).

\bibitem[{\citenamefont{Scalapino}(1995)}]{scalapino95}
\bibinfo{author}{\bibfnamefont{D.}~\bibnamefont{Scalapino}},
  \bibinfo{journal}{Physics Reports} \textbf{\bibinfo{volume}{250}},
  \bibinfo{pages}{329 } (\bibinfo{year}{1995}).

\bibitem[{\citenamefont{Sasmal et~al.}(2008)\citenamefont{Sasmal, Lv, Lorenz,
  Guloy, Chen, Xue, and Chu}}]{sasmal08}
\bibinfo{author}{\bibfnamefont{K.}~\bibnamefont{Sasmal}},
  \bibinfo{author}{\bibfnamefont{B.}~\bibnamefont{Lv}},
  \bibinfo{author}{\bibfnamefont{B.}~\bibnamefont{Lorenz}},
  \bibinfo{author}{\bibfnamefont{A.~M.} \bibnamefont{Guloy}},
  \bibinfo{author}{\bibfnamefont{F.}~\bibnamefont{Chen}},
  \bibinfo{author}{\bibfnamefont{Y.-Y.} \bibnamefont{Xue}}, \bibnamefont{and}
  \bibinfo{author}{\bibfnamefont{C.-W.} \bibnamefont{Chu}},
  \bibinfo{journal}{Phys. Rev. Lett.} \textbf{\bibinfo{volume}{101}},
  \bibinfo{pages}{107007} (\bibinfo{year}{2008}).

\bibitem[{\citenamefont{Dahm et~al.}(2009)\citenamefont{Dahm, Hinkov,
  Borisenko, Kordyuk, Zabolotnyy, Fink, B\"uchner, Scalapino, Hanke, and
  Keimer}}]{dahm09}
\bibinfo{author}{\bibfnamefont{T.}~\bibnamefont{Dahm}},
  \bibinfo{author}{\bibfnamefont{V.}~\bibnamefont{Hinkov}},
  \bibinfo{author}{\bibfnamefont{S.~V.} \bibnamefont{Borisenko}},
  \bibinfo{author}{\bibfnamefont{A.~A.} \bibnamefont{Kordyuk}},
  \bibinfo{author}{\bibfnamefont{V.~B.} \bibnamefont{Zabolotnyy}},
  \bibinfo{author}{\bibfnamefont{J.}~\bibnamefont{Fink}},
  \bibinfo{author}{\bibfnamefont{B.}~\bibnamefont{B\"uchner}},
  \bibinfo{author}{\bibfnamefont{D.~J.} \bibnamefont{Scalapino}},
  \bibinfo{author}{\bibfnamefont{W.}~\bibnamefont{Hanke}}, \bibnamefont{and}
  \bibinfo{author}{\bibfnamefont{B.}~\bibnamefont{Keimer}},
  \bibinfo{journal}{Nat. Phys.} \textbf{\bibinfo{volume}{5}},
  \bibinfo{pages}{217 } (\bibinfo{year}{2009}).

\bibitem[{\citenamefont{Monastra et~al.}(2002)\citenamefont{Monastra, Manghi,
  Rozzi, Arcangeli, Wetli, Neff, Greber, and Osterwalder}}]{monastra02}
\bibinfo{author}{\bibfnamefont{S.}~\bibnamefont{Monastra}},
  \bibinfo{author}{\bibfnamefont{F.}~\bibnamefont{Manghi}},
  \bibinfo{author}{\bibfnamefont{C.~A.} \bibnamefont{Rozzi}},
  \bibinfo{author}{\bibfnamefont{C.}~\bibnamefont{Arcangeli}},
  \bibinfo{author}{\bibfnamefont{E.}~\bibnamefont{Wetli}},
  \bibinfo{author}{\bibfnamefont{H.-J.} \bibnamefont{Neff}},
  \bibinfo{author}{\bibfnamefont{T.}~\bibnamefont{Greber}}, \bibnamefont{and}
  \bibinfo{author}{\bibfnamefont{J.}~\bibnamefont{Osterwalder}},
  \bibinfo{journal}{Phys. Rev. Lett.} \textbf{\bibinfo{volume}{88}},
  \bibinfo{pages}{236402} (\bibinfo{year}{2002}).

\bibitem[{\citenamefont{Knorren et~al.}(2000)\citenamefont{Knorren, Bennemann,
  Burgermeister, and Aeschlimann}}]{knorren00}
\bibinfo{author}{\bibfnamefont{R.}~\bibnamefont{Knorren}},
  \bibinfo{author}{\bibfnamefont{K.~H.} \bibnamefont{Bennemann}},
  \bibinfo{author}{\bibfnamefont{R.}~\bibnamefont{Burgermeister}},
  \bibnamefont{and}
  \bibinfo{author}{\bibfnamefont{M.}~\bibnamefont{Aeschlimann}},
  \bibinfo{journal}{Phys. Rev. B} \textbf{\bibinfo{volume}{61}},
  \bibinfo{pages}{9427} (\bibinfo{year}{2000}).

\bibitem[{\citenamefont{Sch\"afer et~al.}(2004)\citenamefont{Sch\"afer,
  Schrupp, Rotenberg, Rossnagel, Koh, Blaha, and Claessen}}]{schaefer04}
\bibinfo{author}{\bibfnamefont{J.}~\bibnamefont{Sch\"afer}},
  \bibinfo{author}{\bibfnamefont{D.}~\bibnamefont{Schrupp}},
  \bibinfo{author}{\bibfnamefont{E.}~\bibnamefont{Rotenberg}},
  \bibinfo{author}{\bibfnamefont{K.}~\bibnamefont{Rossnagel}},
  \bibinfo{author}{\bibfnamefont{H.}~\bibnamefont{Koh}},
  \bibinfo{author}{\bibfnamefont{P.}~\bibnamefont{Blaha}}, \bibnamefont{and}
  \bibinfo{author}{\bibfnamefont{R.}~\bibnamefont{Claessen}},
  \bibinfo{journal}{Phys. Rev. Lett.} \textbf{\bibinfo{volume}{92}},
  \bibinfo{pages}{097205} (\bibinfo{year}{2004}).

\bibitem[{\citenamefont{Hofmann et~al.}(2009)\citenamefont{Hofmann, Cui,
  Sch\"afer, Meyer, H\"opfner, Blumenstein, Paul, Patthey, Rotenberg,
  B\"unemann et~al.}}]{hofmann09}
\bibinfo{author}{\bibfnamefont{A.}~\bibnamefont{Hofmann}},
  \bibinfo{author}{\bibfnamefont{X.~Y.} \bibnamefont{Cui}},
  \bibinfo{author}{\bibfnamefont{J.}~\bibnamefont{Sch\"afer}},
  \bibinfo{author}{\bibfnamefont{S.}~\bibnamefont{Meyer}},
  \bibinfo{author}{\bibfnamefont{P.}~\bibnamefont{H\"opfner}},
  \bibinfo{author}{\bibfnamefont{C.}~\bibnamefont{Blumenstein}},
  \bibinfo{author}{\bibfnamefont{M.}~\bibnamefont{Paul}},
  \bibinfo{author}{\bibfnamefont{L.}~\bibnamefont{Patthey}},
  \bibinfo{author}{\bibfnamefont{E.}~\bibnamefont{Rotenberg}},
  \bibinfo{author}{\bibfnamefont{J.}~\bibnamefont{B\"unemann}},
  \bibnamefont{et~al.}, \bibinfo{journal}{Phys. Rev. Lett.}
  \textbf{\bibinfo{volume}{102}}, \bibinfo{pages}{187204}
  (\bibinfo{year}{2009}).

\bibitem[{\citenamefont{Mlynczak et~al.}(2019)\citenamefont{Mlynczak,
  M{\"u}ller, Gospodaric, Heider, Aguilera, Bihlmayer, Gehlmann, Jugovac,
  Zamborlini, Tusche et~al.}}]{Mlynczak19}
\bibinfo{author}{\bibfnamefont{E.}~\bibnamefont{Mlynczak}},
  \bibinfo{author}{\bibfnamefont{M.~C. T.~D.} \bibnamefont{M{\"u}ller}},
  \bibinfo{author}{\bibfnamefont{P.}~\bibnamefont{Gospodaric}},
  \bibinfo{author}{\bibfnamefont{T.}~\bibnamefont{Heider}},
  \bibinfo{author}{\bibfnamefont{I.}~\bibnamefont{Aguilera}},
  \bibinfo{author}{\bibfnamefont{G.}~\bibnamefont{Bihlmayer}},
  \bibinfo{author}{\bibfnamefont{M.}~\bibnamefont{Gehlmann}},
  \bibinfo{author}{\bibfnamefont{M.}~\bibnamefont{Jugovac}},
  \bibinfo{author}{\bibfnamefont{G.}~\bibnamefont{Zamborlini}},
  \bibinfo{author}{\bibfnamefont{C.}~\bibnamefont{Tusche}},
  \bibnamefont{et~al.}, \bibinfo{journal}{Nature Communications}
  \textbf{\bibinfo{volume}{10}}, \bibinfo{pages}{505} (\bibinfo{year}{2019}).

\bibitem[{\citenamefont{Georges et~al.}(1996)\citenamefont{Georges, Kotliar,
  Krauth, and Rozenberg}}]{georges96}
\bibinfo{author}{\bibfnamefont{A.}~\bibnamefont{Georges}},
  \bibinfo{author}{\bibfnamefont{G.}~\bibnamefont{Kotliar}},
  \bibinfo{author}{\bibfnamefont{W.}~\bibnamefont{Krauth}}, \bibnamefont{and}
  \bibinfo{author}{\bibfnamefont{M.~J.} \bibnamefont{Rozenberg}},
  \bibinfo{journal}{Rev. Mod. Phys.} \textbf{\bibinfo{volume}{68}},
  \bibinfo{pages}{13} (\bibinfo{year}{1996}).

\bibitem[{\citenamefont{Kotliar et~al.}(2006)\citenamefont{Kotliar, Savrasov,
  Haule, Oudovenko, Parcollet, and Marianetti}}]{kotliar06}
\bibinfo{author}{\bibfnamefont{G.}~\bibnamefont{Kotliar}},
  \bibinfo{author}{\bibfnamefont{S.~Y.} \bibnamefont{Savrasov}},
  \bibinfo{author}{\bibfnamefont{K.}~\bibnamefont{Haule}},
  \bibinfo{author}{\bibfnamefont{V.~S.} \bibnamefont{Oudovenko}},
  \bibinfo{author}{\bibfnamefont{O.}~\bibnamefont{Parcollet}},
  \bibnamefont{and} \bibinfo{author}{\bibfnamefont{C.~A.}
  \bibnamefont{Marianetti}}, \bibinfo{journal}{Rev. Mod. Phys.}
  \textbf{\bibinfo{volume}{78}}, \bibinfo{pages}{865} (\bibinfo{year}{2006}).

\bibitem[{\citenamefont{Katsnelson and Lichtenstein}(1999)}]{katsnelson99}
\bibinfo{author}{\bibfnamefont{M.~I.} \bibnamefont{Katsnelson}}
  \bibnamefont{and} \bibinfo{author}{\bibfnamefont{A.~I.}
  \bibnamefont{Lichtenstein}}, \bibinfo{journal}{Journal of Physics: Condensed
  Matter} \textbf{\bibinfo{volume}{11}}, \bibinfo{pages}{1037}
  (\bibinfo{year}{1999}).

\bibitem[{\citenamefont{Grechnev et~al.}(2007)\citenamefont{Grechnev, Di~Marco,
  Katsnelson, Lichtenstein, Wills, and Eriksson}}]{grechnev07}
\bibinfo{author}{\bibfnamefont{A.}~\bibnamefont{Grechnev}},
  \bibinfo{author}{\bibfnamefont{I.}~\bibnamefont{Di~Marco}},
  \bibinfo{author}{\bibfnamefont{M.~I.} \bibnamefont{Katsnelson}},
  \bibinfo{author}{\bibfnamefont{A.~I.} \bibnamefont{Lichtenstein}},
  \bibinfo{author}{\bibfnamefont{J.}~\bibnamefont{Wills}}, \bibnamefont{and}
  \bibinfo{author}{\bibfnamefont{O.}~\bibnamefont{Eriksson}},
  \bibinfo{journal}{Phys. Rev. B} \textbf{\bibinfo{volume}{76}},
  \bibinfo{pages}{035107} (\bibinfo{year}{2007}).

\bibitem[{\citenamefont{S\'anchez-Barriga
  et~al.}(2009)\citenamefont{S\'anchez-Barriga, Fink, Boni, Di~Marco, Braun,
  Min\'ar, Varykhalov, Rader, Bellini, Manghi et~al.}}]{sanchez-barriga09}
\bibinfo{author}{\bibfnamefont{J.}~\bibnamefont{S\'anchez-Barriga}},
  \bibinfo{author}{\bibfnamefont{J.}~\bibnamefont{Fink}},
  \bibinfo{author}{\bibfnamefont{V.}~\bibnamefont{Boni}},
  \bibinfo{author}{\bibfnamefont{I.}~\bibnamefont{Di~Marco}},
  \bibinfo{author}{\bibfnamefont{J.}~\bibnamefont{Braun}},
  \bibinfo{author}{\bibfnamefont{J.}~\bibnamefont{Min\'ar}},
  \bibinfo{author}{\bibfnamefont{A.}~\bibnamefont{Varykhalov}},
  \bibinfo{author}{\bibfnamefont{O.}~\bibnamefont{Rader}},
  \bibinfo{author}{\bibfnamefont{V.}~\bibnamefont{Bellini}},
  \bibinfo{author}{\bibfnamefont{F.}~\bibnamefont{Manghi}},
  \bibnamefont{et~al.}, \bibinfo{journal}{Phys. Rev. Lett.}
  \textbf{\bibinfo{volume}{103}}, \bibinfo{pages}{267203}
  (\bibinfo{year}{2009}).

\bibitem[{\citenamefont{S\'anchez-Barriga
  et~al.}(2010)\citenamefont{S\'anchez-Barriga, Min\'ar, Braun, Varykhalov,
  Boni, Di~Marco, Rader, Bellini, Manghi, Ebert et~al.}}]{sanchez-barriga10}
\bibinfo{author}{\bibfnamefont{J.}~\bibnamefont{S\'anchez-Barriga}},
  \bibinfo{author}{\bibfnamefont{J.}~\bibnamefont{Min\'ar}},
  \bibinfo{author}{\bibfnamefont{J.}~\bibnamefont{Braun}},
  \bibinfo{author}{\bibfnamefont{A.}~\bibnamefont{Varykhalov}},
  \bibinfo{author}{\bibfnamefont{V.}~\bibnamefont{Boni}},
  \bibinfo{author}{\bibfnamefont{I.}~\bibnamefont{Di~Marco}},
  \bibinfo{author}{\bibfnamefont{O.}~\bibnamefont{Rader}},
  \bibinfo{author}{\bibfnamefont{V.}~\bibnamefont{Bellini}},
  \bibinfo{author}{\bibfnamefont{F.}~\bibnamefont{Manghi}},
  \bibinfo{author}{\bibfnamefont{H.}~\bibnamefont{Ebert}},
  \bibnamefont{et~al.}, \bibinfo{journal}{Phys. Rev. B}
  \textbf{\bibinfo{volume}{82}}, \bibinfo{pages}{104414}
  (\bibinfo{year}{2010}).

\bibitem[{\citenamefont{S\'anchez-Barriga
  et~al.}(2012)\citenamefont{S\'anchez-Barriga, Braun, Min\'ar, Di~Marco,
  Varykhalov, Rader, Boni, Bellini, Manghi, Ebert et~al.}}]{sanchez-barriga12}
\bibinfo{author}{\bibfnamefont{J.}~\bibnamefont{S\'anchez-Barriga}},
  \bibinfo{author}{\bibfnamefont{J.}~\bibnamefont{Braun}},
  \bibinfo{author}{\bibfnamefont{J.}~\bibnamefont{Min\'ar}},
  \bibinfo{author}{\bibfnamefont{I.}~\bibnamefont{Di~Marco}},
  \bibinfo{author}{\bibfnamefont{A.}~\bibnamefont{Varykhalov}},
  \bibinfo{author}{\bibfnamefont{O.}~\bibnamefont{Rader}},
  \bibinfo{author}{\bibfnamefont{V.}~\bibnamefont{Boni}},
  \bibinfo{author}{\bibfnamefont{V.}~\bibnamefont{Bellini}},
  \bibinfo{author}{\bibfnamefont{F.}~\bibnamefont{Manghi}},
  \bibinfo{author}{\bibfnamefont{H.}~\bibnamefont{Ebert}},
  \bibnamefont{et~al.}, \bibinfo{journal}{Phys. Rev. B}
  \textbf{\bibinfo{volume}{85}}, \bibinfo{pages}{205109}
  (\bibinfo{year}{2012}).

\bibitem[{\citenamefont{Onida et~al.}(2002)\citenamefont{Onida, Reining, and
  Rubio}}]{onida02}
\bibinfo{author}{\bibfnamefont{G.}~\bibnamefont{Onida}},
  \bibinfo{author}{\bibfnamefont{L.}~\bibnamefont{Reining}}, \bibnamefont{and}
  \bibinfo{author}{\bibfnamefont{A.}~\bibnamefont{Rubio}},
  \bibinfo{journal}{Rev. Mod. Phys.} \textbf{\bibinfo{volume}{74}},
  \bibinfo{pages}{601} (\bibinfo{year}{2002}).

\bibitem[{\citenamefont{Aulbur et~al.}(1999)\citenamefont{Aulbur, J\"onsson,
  and Wilkins}}]{Aulbur99}
\bibinfo{author}{\bibfnamefont{W.~G.} \bibnamefont{Aulbur}},
  \bibinfo{author}{\bibfnamefont{L.}~\bibnamefont{J\"onsson}},
  \bibnamefont{and} \bibinfo{author}{\bibfnamefont{J.~W.}
  \bibnamefont{Wilkins}} (\bibinfo{publisher}{Academic Press},
  \bibinfo{address}{Cambridge MA}, \bibinfo{year}{1999}),
  vol.~\bibinfo{volume}{54} of \emph{\bibinfo{series}{Solid State Physics}},
  pp. \bibinfo{pages}{1 -- 218}.

\bibitem[{\citenamefont{Yamasaki and Fujiwara}(2003)}]{yamasaki03}
\bibinfo{author}{\bibfnamefont{A.}~\bibnamefont{Yamasaki}} \bibnamefont{and}
  \bibinfo{author}{\bibfnamefont{T.}~\bibnamefont{Fujiwara}},
  \bibinfo{journal}{Journal of the Physical Society of Japan}
  \textbf{\bibinfo{volume}{72}}, \bibinfo{pages}{607} (\bibinfo{year}{2003}).

\bibitem[{\citenamefont{Aryasetiawan}(1992)}]{aryasetiawan92}
\bibinfo{author}{\bibfnamefont{F.}~\bibnamefont{Aryasetiawan}},
  \bibinfo{journal}{Phys. Rev. B} \textbf{\bibinfo{volume}{46}},
  \bibinfo{pages}{13051} (\bibinfo{year}{1992}).

\bibitem[{\citenamefont{Friedrich et~al.}(2010)\citenamefont{Friedrich,
  Bl\"ugel, and Schindlmayr}}]{friedrich10}
\bibinfo{author}{\bibfnamefont{C.}~\bibnamefont{Friedrich}},
  \bibinfo{author}{\bibfnamefont{S.}~\bibnamefont{Bl\"ugel}}, \bibnamefont{and}
  \bibinfo{author}{\bibfnamefont{A.}~\bibnamefont{Schindlmayr}},
  \bibinfo{journal}{Phys. Rev. B} \textbf{\bibinfo{volume}{81}},
  \bibinfo{pages}{125102} (\bibinfo{year}{2010}).

\bibitem[{\citenamefont{Eastman et~al.}(1980)\citenamefont{Eastman, Himpsel,
  and Knapp}}]{eastman80}
\bibinfo{author}{\bibfnamefont{D.~E.} \bibnamefont{Eastman}},
  \bibinfo{author}{\bibfnamefont{F.~J.} \bibnamefont{Himpsel}},
  \bibnamefont{and} \bibinfo{author}{\bibfnamefont{J.~A.} \bibnamefont{Knapp}},
  \bibinfo{journal}{Phys. Rev. Lett.} \textbf{\bibinfo{volume}{44}},
  \bibinfo{pages}{95} (\bibinfo{year}{1980}).

\bibitem[{\citenamefont{H{\"o}chst et~al.}(1977)\citenamefont{H{\"o}chst,
  H{\"u}fner, and Goldmann}}]{hoechst77}
\bibinfo{author}{\bibfnamefont{H.}~\bibnamefont{H{\"o}chst}},
  \bibinfo{author}{\bibfnamefont{S.}~\bibnamefont{H{\"u}fner}},
  \bibnamefont{and} \bibinfo{author}{\bibfnamefont{A.}~\bibnamefont{Goldmann}},
  \bibinfo{journal}{Zeitschrift f{\"u}r Physik B Condensed Matter}
  \textbf{\bibinfo{volume}{26}}, \bibinfo{pages}{133} (\bibinfo{year}{1977}).

\bibitem[{\citenamefont{Himpsel et~al.}(1979)\citenamefont{Himpsel, Knapp, and
  Eastman}}]{himpsel79}
\bibinfo{author}{\bibfnamefont{F.~J.} \bibnamefont{Himpsel}},
  \bibinfo{author}{\bibfnamefont{J.~A.} \bibnamefont{Knapp}}, \bibnamefont{and}
  \bibinfo{author}{\bibfnamefont{D.~E.} \bibnamefont{Eastman}},
  \bibinfo{journal}{Phys. Rev. B} \textbf{\bibinfo{volume}{19}},
  \bibinfo{pages}{2919} (\bibinfo{year}{1979}).

\bibitem[{\citenamefont{Eberhardt and Plummer}(1980)}]{eberhardt80}
\bibinfo{author}{\bibfnamefont{W.}~\bibnamefont{Eberhardt}} \bibnamefont{and}
  \bibinfo{author}{\bibfnamefont{E.~W.} \bibnamefont{Plummer}},
  \bibinfo{journal}{Phys. Rev. B} \textbf{\bibinfo{volume}{21}},
  \bibinfo{pages}{3245} (\bibinfo{year}{1980}).

\bibitem[{\citenamefont{Heimann et~al.}(1981)\citenamefont{Heimann, Himpsel,
  and Eastman}}]{heimann81}
\bibinfo{author}{\bibfnamefont{P.}~\bibnamefont{Heimann}},
  \bibinfo{author}{\bibfnamefont{F.}~\bibnamefont{Himpsel}}, \bibnamefont{and}
  \bibinfo{author}{\bibfnamefont{D.}~\bibnamefont{Eastman}},
  \bibinfo{journal}{Solid State Communications} \textbf{\bibinfo{volume}{39}},
  \bibinfo{pages}{219 } (\bibinfo{year}{1981}).

\bibitem[{\citenamefont{Kirby et~al.}(1985)\citenamefont{Kirby, Kisker, King,
  and Garwin}}]{kirby85}
\bibinfo{author}{\bibfnamefont{R.}~\bibnamefont{Kirby}},
  \bibinfo{author}{\bibfnamefont{E.}~\bibnamefont{Kisker}},
  \bibinfo{author}{\bibfnamefont{F.}~\bibnamefont{King}}, \bibnamefont{and}
  \bibinfo{author}{\bibfnamefont{E.}~\bibnamefont{Garwin}},
  \bibinfo{journal}{Solid State Communications} \textbf{\bibinfo{volume}{56}},
  \bibinfo{pages}{425 } (\bibinfo{year}{1985}).

\bibitem[{\citenamefont{H{\"o}chst
  et~al.}(1976{\natexlab{a}})\citenamefont{H{\"o}chst, Goldmann, and
  H{\"u}fner}}]{hoechst76}
\bibinfo{author}{\bibfnamefont{H.}~\bibnamefont{H{\"o}chst}},
  \bibinfo{author}{\bibfnamefont{A.}~\bibnamefont{Goldmann}}, \bibnamefont{and}
  \bibinfo{author}{\bibfnamefont{S.}~\bibnamefont{H{\"u}fner}},
  \bibinfo{journal}{Zeitschrift f{\"u}r Physik B Condensed Matter}
  \textbf{\bibinfo{volume}{24}}, \bibinfo{pages}{245}
  (\bibinfo{year}{1976}{\natexlab{a}}).

\bibitem[{\citenamefont{H{\"o}chst
  et~al.}(1976{\natexlab{b}})\citenamefont{H{\"o}chst, H{\"u}fner, and
  Goldmann}}]{hoechst76a}
\bibinfo{author}{\bibfnamefont{H.}~\bibnamefont{H{\"o}chst}},
  \bibinfo{author}{\bibfnamefont{S.}~\bibnamefont{H{\"u}fner}},
  \bibnamefont{and} \bibinfo{author}{\bibfnamefont{A.}~\bibnamefont{Goldmann}},
  \bibinfo{journal}{Physics Letters A} \textbf{\bibinfo{volume}{57}},
  \bibinfo{pages}{265 } (\bibinfo{year}{1976}{\natexlab{b}}).

\bibitem[{\citenamefont{Hedin}(1965)}]{hedin65}
\bibinfo{author}{\bibfnamefont{L.}~\bibnamefont{Hedin}},
  \bibinfo{journal}{Phys. Rev.} \textbf{\bibinfo{volume}{139}},
  \bibinfo{pages}{A796} (\bibinfo{year}{1965}).

\bibitem[{\citenamefont{Springer et~al.}(1998)\citenamefont{Springer,
  Aryasetiawan, and Karlsson}}]{springer98}
\bibinfo{author}{\bibfnamefont{M.}~\bibnamefont{Springer}},
  \bibinfo{author}{\bibfnamefont{F.}~\bibnamefont{Aryasetiawan}},
  \bibnamefont{and} \bibinfo{author}{\bibfnamefont{K.}~\bibnamefont{Karlsson}},
  \bibinfo{journal}{Phys. Rev. Lett.} \textbf{\bibinfo{volume}{80}},
  \bibinfo{pages}{2389} (\bibinfo{year}{1998}).

\bibitem[{\citenamefont{Zhukov et~al.}(2004)\citenamefont{Zhukov, Chulkov, and
  Echenique}}]{zhukov04}
\bibinfo{author}{\bibfnamefont{V.~P.} \bibnamefont{Zhukov}},
  \bibinfo{author}{\bibfnamefont{E.~V.} \bibnamefont{Chulkov}},
  \bibnamefont{and} \bibinfo{author}{\bibfnamefont{P.~M.}
  \bibnamefont{Echenique}}, \bibinfo{journal}{Phys. Rev. Lett.}
  \textbf{\bibinfo{volume}{93}}, \bibinfo{pages}{096401}
  (\bibinfo{year}{2004}).

\bibitem[{\citenamefont{Romaniello et~al.}(2012)\citenamefont{Romaniello,
  Bechstedt, and Reining}}]{romaniello12}
\bibinfo{author}{\bibfnamefont{P.}~\bibnamefont{Romaniello}},
  \bibinfo{author}{\bibfnamefont{F.}~\bibnamefont{Bechstedt}},
  \bibnamefont{and} \bibinfo{author}{\bibfnamefont{L.}~\bibnamefont{Reining}},
  \bibinfo{journal}{Phys. Rev. B} \textbf{\bibinfo{volume}{85}},
  \bibinfo{pages}{155131} (\bibinfo{year}{2012}).

\bibitem[{\citenamefont{Aryasetiawan and Karlsson}(1999)}]{aryasetiawan99}
\bibinfo{author}{\bibfnamefont{F.}~\bibnamefont{Aryasetiawan}}
  \bibnamefont{and} \bibinfo{author}{\bibfnamefont{K.}~\bibnamefont{Karlsson}},
  \bibinfo{journal}{Phys. Rev. B} \textbf{\bibinfo{volume}{60}},
  \bibinfo{pages}{7419} (\bibinfo{year}{1999}).

\bibitem[{\citenamefont{\ifmmode \mbox{\c{S}}\else \c{S}\fi{}a\ifmmode
  \mbox{\c{s}}\else \c{s}\fi{}\ifmmode \imath \else \i
  \fi{}o\ifmmode~\breve{g}\else \u{g}\fi{}lu
  et~al.}(2010)\citenamefont{\ifmmode \mbox{\c{S}}\else \c{S}\fi{}a\ifmmode
  \mbox{\c{s}}\else \c{s}\fi{}\ifmmode \imath \else \i
  \fi{}o\ifmmode~\breve{g}\else \u{g}\fi{}lu, Schindlmayr, Friedrich, Freimuth,
  and Bl\"ugel}}]{sasioglu10}
\bibinfo{author}{\bibfnamefont{E.}~\bibnamefont{\ifmmode \mbox{\c{S}}\else
  \c{S}\fi{}a\ifmmode \mbox{\c{s}}\else \c{s}\fi{}\ifmmode \imath \else \i
  \fi{}o\ifmmode~\breve{g}\else \u{g}\fi{}lu}},
  \bibinfo{author}{\bibfnamefont{A.}~\bibnamefont{Schindlmayr}},
  \bibinfo{author}{\bibfnamefont{C.}~\bibnamefont{Friedrich}},
  \bibinfo{author}{\bibfnamefont{F.}~\bibnamefont{Freimuth}}, \bibnamefont{and}
  \bibinfo{author}{\bibfnamefont{S.}~\bibnamefont{Bl\"ugel}},
  \bibinfo{journal}{Phys. Rev. B} \textbf{\bibinfo{volume}{81}},
  \bibinfo{pages}{054434} (\bibinfo{year}{2010}).

\bibitem[{\citenamefont{Friedrich et~al.}(2014)\citenamefont{Friedrich,
  \ifmmode \mbox{\c{S}}\else \c{S}\fi{}a\ifmmode \mbox{\c{s}}\else
  \c{s}\fi{}\ifmmode \imath \else \i \fi{}o\ifmmode~\breve{g}\else
  \u{g}\fi{}lu, M\"uller, Schindlmayr, and Bl\"ugel}}]{friedrich14}
\bibinfo{author}{\bibfnamefont{C.}~\bibnamefont{Friedrich}},
  \bibinfo{author}{\bibfnamefont{E.}~\bibnamefont{\ifmmode \mbox{\c{S}}\else
  \c{S}\fi{}a\ifmmode \mbox{\c{s}}\else \c{s}\fi{}\ifmmode \imath \else \i
  \fi{}o\ifmmode~\breve{g}\else \u{g}\fi{}lu}},
  \bibinfo{author}{\bibfnamefont{M.}~\bibnamefont{M\"uller}},
  \bibinfo{author}{\bibfnamefont{A.}~\bibnamefont{Schindlmayr}},
  \bibnamefont{and} \bibinfo{author}{\bibfnamefont{S.}~\bibnamefont{Bl\"ugel}},
  in \emph{\bibinfo{booktitle}{First Principles Approaches to Spectroscopic
  Properties of Complex Materials}}, edited by
  \bibinfo{editor}{\bibfnamefont{C.}~\bibnamefont{Di~Valentin}},
  \bibinfo{editor}{\bibfnamefont{S.}~\bibnamefont{Botti}}, \bibnamefont{and}
  \bibinfo{editor}{\bibfnamefont{M.}~\bibnamefont{Cococcioni}}
  (\bibinfo{publisher}{Springer Berlin Heidelberg}, \bibinfo{year}{2014}), vol.
  \bibinfo{volume}{347} of \emph{\bibinfo{series}{Topics in Current
  Chemistry}}, pp. \bibinfo{pages}{259--301}.

\bibitem[{\citenamefont{Friedrich et~al.}(2018)\citenamefont{Friedrich,
  M\"uller, and Bl\"ugel}}]{friedrich18}
\bibinfo{author}{\bibfnamefont{C.}~\bibnamefont{Friedrich}},
  \bibinfo{author}{\bibfnamefont{M.~C. T.~D.} \bibnamefont{M\"uller}},
  \bibnamefont{and} \bibinfo{author}{\bibfnamefont{S.}~\bibnamefont{Bl\"ugel}},
  in \emph{\bibinfo{booktitle}{Handbook of Materials Modeling. Volume 1
  Methods: Theory and Modeling}}, edited by
  \bibinfo{editor}{\bibfnamefont{S.}~\bibnamefont{Yip}} \bibnamefont{and}
  \bibinfo{editor}{\bibfnamefont{W.}~\bibnamefont{Andreoni}}
  (\bibinfo{publisher}{Springer Berlin Heidelberg}, \bibinfo{year}{2018}).

\bibitem[{\citenamefont{M\"uller et~al.}(2016)\citenamefont{M\"uller,
  Friedrich, and Bl\"ugel}}]{mueller16}
\bibinfo{author}{\bibfnamefont{M.~C. T.~D.} \bibnamefont{M\"uller}},
  \bibinfo{author}{\bibfnamefont{C.}~\bibnamefont{Friedrich}},
  \bibnamefont{and} \bibinfo{author}{\bibfnamefont{S.}~\bibnamefont{Bl\"ugel}},
  \bibinfo{journal}{Phys. Rev. B} \textbf{\bibinfo{volume}{94}},
  \bibinfo{pages}{064433} (\bibinfo{year}{2016}).

\bibitem[{\citenamefont{Moriya}(1985)}]{moriya85}
\bibinfo{author}{\bibfnamefont{T.}~\bibnamefont{Moriya}},
  \emph{\bibinfo{title}{Spin Fluctuations in Itinerant Electron Magnetism}}
  (\bibinfo{publisher}{Springer Berlin Heidelberg}, \bibinfo{year}{1985}).

\bibitem[{\citenamefont{Nambu}(1960)}]{nambu60}
\bibinfo{author}{\bibfnamefont{Y.}~\bibnamefont{Nambu}},
  \bibinfo{journal}{Phys. Rev.} \textbf{\bibinfo{volume}{117}},
  \bibinfo{pages}{648} (\bibinfo{year}{1960}).

\bibitem[{\citenamefont{Goldstone}(1961)}]{goldstone61}
\bibinfo{author}{\bibfnamefont{J.}~\bibnamefont{Goldstone}},
  \bibinfo{journal}{Il Nuovo Cimento} \textbf{\bibinfo{volume}{19}},
  \bibinfo{pages}{154} (\bibinfo{year}{1961}).

\bibitem[{fle()}]{fleur}
\url{http://www.flapw.de}.

\bibitem[{\citenamefont{Freimuth et~al.}(2008)\citenamefont{Freimuth,
  Mokrousov, Wortmann, Heinze, and Bl\"ugel}}]{freimuth08}
\bibinfo{author}{\bibfnamefont{F.}~\bibnamefont{Freimuth}},
  \bibinfo{author}{\bibfnamefont{Y.}~\bibnamefont{Mokrousov}},
  \bibinfo{author}{\bibfnamefont{D.}~\bibnamefont{Wortmann}},
  \bibinfo{author}{\bibfnamefont{S.}~\bibnamefont{Heinze}}, \bibnamefont{and}
  \bibinfo{author}{\bibfnamefont{S.}~\bibnamefont{Bl\"ugel}},
  \bibinfo{journal}{Phys. Rev. B} \textbf{\bibinfo{volume}{78}},
  \bibinfo{pages}{035120} (\bibinfo{year}{2008}).

\bibitem[{\citenamefont{Rojas et~al.}(1995)\citenamefont{Rojas, Godby, and
  Needs}}]{rojas95}
\bibinfo{author}{\bibfnamefont{H.~N.} \bibnamefont{Rojas}},
  \bibinfo{author}{\bibfnamefont{R.~W.} \bibnamefont{Godby}}, \bibnamefont{and}
  \bibinfo{author}{\bibfnamefont{R.~J.} \bibnamefont{Needs}},
  \bibinfo{journal}{Phys. Rev. Lett.} \textbf{\bibinfo{volume}{74}},
  \bibinfo{pages}{1827} (\bibinfo{year}{1995}).

\bibitem[{\citenamefont{Rieger et~al.}(1999)\citenamefont{Rieger, Steinbeck,
  White, Rojas, and Godby}}]{rieger99}
\bibinfo{author}{\bibfnamefont{M.~M.} \bibnamefont{Rieger}},
  \bibinfo{author}{\bibfnamefont{L.}~\bibnamefont{Steinbeck}},
  \bibinfo{author}{\bibfnamefont{I.}~\bibnamefont{White}},
  \bibinfo{author}{\bibfnamefont{H.}~\bibnamefont{Rojas}}, \bibnamefont{and}
  \bibinfo{author}{\bibfnamefont{R.}~\bibnamefont{Godby}},
  \bibinfo{journal}{Computer Physics Communications}
  \textbf{\bibinfo{volume}{117}}, \bibinfo{pages}{211 } (\bibinfo{year}{1999}).

\bibitem[{\citenamefont{Baker and Graves-Morris}(2010)}]{baker10}
\bibinfo{author}{\bibfnamefont{G.}~\bibnamefont{Baker}} \bibnamefont{and}
  \bibinfo{author}{\bibfnamefont{P.}~\bibnamefont{Graves-Morris}},
  \emph{\bibinfo{title}{Pad{\'e} Approximants}}, Encyclopedia of Mathematics
  and its Applications (\bibinfo{publisher}{Cambridge University Press},
  \bibinfo{address}{Cambridge UK}, \bibinfo{year}{2010}).

\bibitem[{\citenamefont{Friedrich}(2019)}]{friedrich19}
\bibinfo{author}{\bibfnamefont{C.}~\bibnamefont{Friedrich}}
  (\bibinfo{year}{2019}), \bibinfo{note}{submitted}.

\bibitem[{\citenamefont{Schindlmayr}(1997)}]{schindlmayr97}
\bibinfo{author}{\bibfnamefont{A.}~\bibnamefont{Schindlmayr}},
  \bibinfo{journal}{Phys. Rev. B} \textbf{\bibinfo{volume}{56}},
  \bibinfo{pages}{3528} (\bibinfo{year}{1997}).

\bibitem[{\citenamefont{Pollehn et~al.}(1998)\citenamefont{Pollehn,
  Schindlmayr, and Godby}}]{pollehn98}
\bibinfo{author}{\bibfnamefont{T.~J.} \bibnamefont{Pollehn}},
  \bibinfo{author}{\bibfnamefont{A.}~\bibnamefont{Schindlmayr}},
  \bibnamefont{and} \bibinfo{author}{\bibfnamefont{R.~W.} \bibnamefont{Godby}},
  \bibinfo{journal}{Journal of Physics: Condensed Matter}
  \textbf{\bibinfo{volume}{10}}, \bibinfo{pages}{1273} (\bibinfo{year}{1998}).

\bibitem[{\citenamefont{Fetter and Walecka}(1971)}]{FetterWalecka71}
\bibinfo{author}{\bibfnamefont{A.~L.} \bibnamefont{Fetter}} \bibnamefont{and}
  \bibinfo{author}{\bibfnamefont{J.~D.} \bibnamefont{Walecka}},
  \emph{\bibinfo{title}{Quantum theory of many-particle systems}}
  (\bibinfo{publisher}{McGraw-Hill}, \bibinfo{address}{New York},
  \bibinfo{year}{1971}).

\bibitem[{\citenamefont{Krause and Th\"ornig}(2019)}]{JSC}
\bibinfo{author}{\bibfnamefont{D.}~\bibnamefont{Krause}} \bibnamefont{and}
  \bibinfo{author}{\bibfnamefont{P.}~\bibnamefont{Th\"ornig}},
  \bibinfo{journal}{Journal of large-scale research facilities}
  \textbf{\bibinfo{volume}{4}}, \bibinfo{pages}{A132} (\bibinfo{year}{2019}).

\bibitem[{\citenamefont{Hong and Mills}(1999)}]{mills99}
\bibinfo{author}{\bibfnamefont{J.}~\bibnamefont{Hong}} \bibnamefont{and}
  \bibinfo{author}{\bibfnamefont{D.~L.} \bibnamefont{Mills}},
  \bibinfo{journal}{Phys. Rev. B} \textbf{\bibinfo{volume}{59}},
  \bibinfo{pages}{13840} (\bibinfo{year}{1999}).

\bibitem[{\citenamefont{Cui et~al.}(2007)\citenamefont{Cui, Shimada, Hoesch,
  Sakisaka, Kato, Aiura, Higashiguchi, Miura, Namatame, and Taniguchi}}]{cui07}
\bibinfo{author}{\bibfnamefont{X.}~\bibnamefont{Cui}},
  \bibinfo{author}{\bibfnamefont{K.}~\bibnamefont{Shimada}},
  \bibinfo{author}{\bibfnamefont{M.}~\bibnamefont{Hoesch}},
  \bibinfo{author}{\bibfnamefont{Y.}~\bibnamefont{Sakisaka}},
  \bibinfo{author}{\bibfnamefont{H.}~\bibnamefont{Kato}},
  \bibinfo{author}{\bibfnamefont{Y.}~\bibnamefont{Aiura}},
  \bibinfo{author}{\bibfnamefont{M.}~\bibnamefont{Higashiguchi}},
  \bibinfo{author}{\bibfnamefont{Y.}~\bibnamefont{Miura}},
  \bibinfo{author}{\bibfnamefont{H.}~\bibnamefont{Namatame}}, \bibnamefont{and}
  \bibinfo{author}{\bibfnamefont{M.}~\bibnamefont{Taniguchi}},
  \bibinfo{journal}{Surface Science} \textbf{\bibinfo{volume}{601}},
  \bibinfo{pages}{4010 } (\bibinfo{year}{2007}), \bibinfo{note}{eCOSS-24}.

\bibitem[{\citenamefont{Cui et~al.}(2010)\citenamefont{Cui, Shimada, Sakisaka,
  Kato, Hoesch, Oguchi, Aiura, Namatame, and Taniguchi}}]{cui10}
\bibinfo{author}{\bibfnamefont{X.~Y.} \bibnamefont{Cui}},
  \bibinfo{author}{\bibfnamefont{K.}~\bibnamefont{Shimada}},
  \bibinfo{author}{\bibfnamefont{Y.}~\bibnamefont{Sakisaka}},
  \bibinfo{author}{\bibfnamefont{H.}~\bibnamefont{Kato}},
  \bibinfo{author}{\bibfnamefont{M.}~\bibnamefont{Hoesch}},
  \bibinfo{author}{\bibfnamefont{T.}~\bibnamefont{Oguchi}},
  \bibinfo{author}{\bibfnamefont{Y.}~\bibnamefont{Aiura}},
  \bibinfo{author}{\bibfnamefont{H.}~\bibnamefont{Namatame}}, \bibnamefont{and}
  \bibinfo{author}{\bibfnamefont{M.}~\bibnamefont{Taniguchi}},
  \bibinfo{journal}{Phys. Rev. B} \textbf{\bibinfo{volume}{82}},
  \bibinfo{pages}{195132} (\bibinfo{year}{2010}).

\bibitem[{\citenamefont{Mahan}(2000)}]{mahan00}
\bibinfo{author}{\bibfnamefont{G.~D.} \bibnamefont{Mahan}},
  \emph{\bibinfo{title}{Many-Particle Physics}} (\bibinfo{publisher}{Springer
  US}, \bibinfo{year}{2000}).

\end{thebibliography}

\end{document}